\documentclass[twoside,twocolumn,9pt]{article}
\usepackage{extsizes}
\usepackage[super,sort&compress,comma]{natbib} 
\usepackage[version=3]{mhchem}
\usepackage[left=1.5cm, right=1.5cm, top=1.785cm, bottom=2.0cm]{geometry}
\usepackage{balance}
\usepackage{mathptmx}
\usepackage{sectsty}
\usepackage{graphicx} 
\usepackage{lastpage}
\usepackage[format=plain,justification=justified,singlelinecheck=false,font={stretch=1.125,small,sf},labelfont=bf,labelsep=space]{caption}
\usepackage{float}
\usepackage{fancyhdr}
\usepackage{fnpos}
\usepackage[english]{babel}
\addto{\captionsenglish}{%
  
}
\usepackage{array}
\usepackage{droidsans}
\usepackage{charter}
\usepackage[T1]{fontenc}
\usepackage[usenames,dvipsnames]{xcolor}
\usepackage{setspace}
\usepackage[compact]{titlesec}
\usepackage{hyperref}

\definecolor{cream}{RGB}{222,217,201}

\newcommand{\nvec}{\hat{\bf n}}

\newcommand{\rvec}{\hat{\bf r}}
\usepackage{amsmath}
\usepackage{graphicx}
\usepackage{bm}
\usepackage{xcolor}
\usepackage{authblk}
\usepackage{lineno}
\usepackage{tabularx,ragged2e} 
\usepackage{amssymb} 
\newcolumntype{L}{>{\RaggedRight\arraybackslash}X} 
\usepackage{enumitem}

\graphicspath{{./Figure_main/}}

    \usepackage{xr}
    \makeatletter
    
    \newcommand*{\addFileDependency}[1]{
    \typeout{(#1)}
    \@addtofilelist{#1}
    \IfFileExists{#1}{}{\typeout{No file #1.}}
    }\makeatother
    
    \newcommand*{\myexternaldocument}[1]{%
    \externaldocument{#1}%
    \addFileDependency{#1.tex}%
    \addFileDependency{#1.aux}%
    }
    \myexternaldocument{2_SI}

\DeclareMathOperator*{\argmax}{arg\,max}
\def\NY#1{\textcolor{Black}{{#1}}} 
\def\UL#1{\textcolor{Black}{{#1}}} 

\begin{document}

\pagestyle{fancy}
\thispagestyle{plain}
\fancypagestyle{plain}{
\renewcommand{\headrulewidth}{0pt}
}

\makeFNbottom
\makeatletter
\renewcommand\LARGE{\@setfontsize\LARGE{15pt}{17}}
\renewcommand\Large{\@setfontsize\Large{12pt}{14}}
\renewcommand\large{\@setfontsize\large{10pt}{12}}
\renewcommand\footnotesize{\@setfontsize\footnotesize{7pt}{10}}
\makeatother

\renewcommand{\thefootnote}{\fnsymbol{footnote}}
\renewcommand\footnoterule{\vspace*{1pt}%
\color{cream}\hrule width 3.5in height 0.4pt \color{black}\vspace*{5pt}} 
\setcounter{secnumdepth}{5}

\makeatletter 
\renewcommand\@biblabel[1]{#1}            
\renewcommand\@makefntext[1]%
{\noindent\makebox[0pt][r]{\@thefnmark\,}#1}
\makeatother 
\renewcommand{\figurename}{\small{Fig.}~}
\sectionfont{\sffamily\Large}
\subsectionfont{\normalsize}
\subsubsectionfont{\bf}
\setstretch{1.125} 
\setlength{\skip\footins}{0.8cm}
\setlength{\footnotesep}{0.25cm}
\setlength{\jot}{10pt}
\titlespacing*{\section}{0pt}{4pt}{4pt}
\titlespacing*{\subsection}{0pt}{15pt}{1pt}

\fancyfoot{}
\fancyfoot[LO,RE]{\vspace{-7.1pt}\includegraphics[height=9pt]{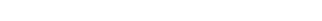}}
\fancyfoot[CO]{\vspace{-7.1pt}\hspace{13.2cm}\includegraphics{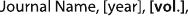}}
\fancyfoot[CE]{\vspace{-7.2pt}\hspace{-14.2cm}\includegraphics{head_foot/RF}}
\fancyfoot[RO]{\footnotesize{\sffamily{1--\pageref{LastPage} ~\textbar  \hspace{2pt}\thepage}}}
\fancyfoot[LE]{\footnotesize{\sffamily{\thepage~\textbar\hspace{3.45cm} 1--\pageref{LastPage}}}}
\fancyhead{}
\renewcommand{\headrulewidth}{0pt} 
\renewcommand{\footrulewidth}{0pt}
\setlength{\arrayrulewidth}{1pt}
\setlength{\columnsep}{6.5mm}
\setlength\bibsep{1pt}

\makeatletter 
\newlength{\figrulesep} 
\setlength{\figrulesep}{0.5\textfloatsep} 

\newcommand{\topfigrule}{\vspace*{-1pt}%
\noindent{\color{cream}\rule[-\figrulesep]{\columnwidth}{1.5pt}} }

\newcommand{\botfigrule}{\vspace*{-2pt}%
\noindent{\color{cream}\rule[\figrulesep]{\columnwidth}{1.5pt}} }

\newcommand{\dblfigrule}{\vspace*{-1pt}%
\noindent{\color{cream}\rule[-\figrulesep]{\textwidth}{1.5pt}} }

\makeatother

\twocolumn[
  \begin{@twocolumnfalse}
{\includegraphics[height=30pt]{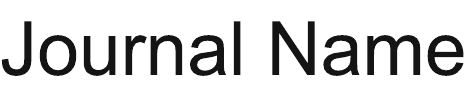}\hfill\raisebox{0pt}[0pt][0pt]{\includegraphics[height=55pt]{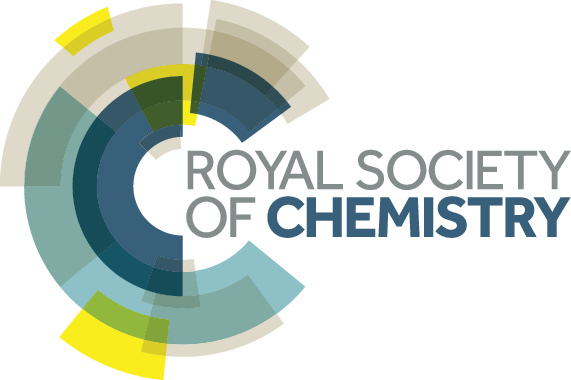}}\\[1ex]
\includegraphics[width=18.5cm]{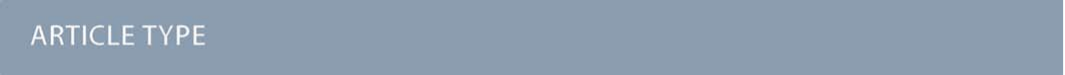}}\par
\vspace{1em}
\sffamily
\begin{tabular}{m{4.5cm} p{13.5cm} }

\includegraphics{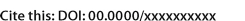} & \noindent\LARGE{\textbf{Dynamic control of self-assembly of quasicrystalline structures through reinforcement learning}} \\
\vspace{0.3cm} & \vspace{0.3cm} \\

 & \noindent\large{Uyen Tu Lieu,$^{\ast}$\text{$^{\dag a,b}$} Natsuhiko Yoshinaga\textit{$^{\dag a,b}$}} \\

\includegraphics{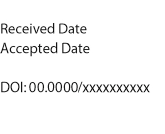} & \noindent\normalsize{
We propose reinforcement learning to control the dynamical self-assembly of the dodecagonal quasicrystal (DDQC) from patchy particles. 
The patchy particles have anisotropic interactions with other particles and form DDQC. 
However, their structures at steady states are significantly influenced by the kinetic pathways of their structural formation. We estimate the best policy of temperature control trained by the Q-learning method and demonstrate that we can generate DDQC with few defects using the estimated policy.
\UL{It is} found that reinforcement learning autonomously discovers \UL{a characteristic} temperature at which structural fluctuations enhance the chance of forming a globally stable state. 
The estimated policy guides the system toward the \UL{characteristic} temperature to assist the formation of DDQC. 
\UL{We also illustrate the performance of RL when the target is metastable or unstable.}
To clarify the success of the learning, we analyse a simple model describing the kinetics of structural changes through the motion in a triple-well potential. 
} \\

\end{tabular}

 \end{@twocolumnfalse} \vspace{0.6cm}

  ]

\renewcommand*\rmdefault{bch}\normalfont\upshape
\rmfamily
\section*{}
\vspace{-1cm}

\footnotetext{\textit{$^{a}$~Future University Hakodate, Kamedanakano-cho 116-2, Hokkaido 041-8655, Japan. E-mail: uyenlieu@fun.ac.jp}}
\footnotetext{\textit{$^{b}$~Mathematics for Advanced Materials-OIL, AIST, Katahira 2-1-1, Sendai 980-8577, Japan. Tel: +81-(0)22-237-8017; E-mail: yoshinaga@fun.ac.jp}}


\footnotetext{\dag~These authors contributed equally to this work.}


\section{Introduction}

Nano- and colloidal self-assembly is promising due to its high potential in creating complex structures with emergent photonic,\cite{Hynninen:2007,He:2020} magnetic,\cite{Tamura:2021} and electronic\cite{Deguchi:2012} properties.
To make various self-assembly structures, several methods have been proposed, such as patchy particles,\cite{Chen:2011a,venturarosales_2020,lieu_2022b} non-spherical particles,\cite{Glotzer:2007} and particles with non-monotonic interactions.\cite{Engel:2015}
Among those, the patchy particle, which has anisotropic interactions, is a good candidate due to its high flexibility in designing the interactions and the capability to form complex structures.\cite{venturarosales_2020,geng_2021} 
In fact, complex structures, such as diamonds and quasicrystals, are reproduced by using patchy particles.
Still, designing a desired structure remains a formidable task and relies on trial and error.

Recently, there has been growing interest in the inverse design of desired self-assembly structures. 
In the conventional forward-type approach, we start from a given model with a specific type of interaction between particles and tune its parameters to analyse the obtained structure.
In contrast, the inverse design estimates the model from the desired structure.
This approach has been successfully applied to several complex structures, such as quasicrystals.
\cite{Kumar:2019,Ma:2019,lieu_2022a,Yoshinaga:2020}
However, so far, most of the methods of the inverse design rely on static control, such as optimisation of parameters in the potential interactions, and  do not take into account the kinetic process of self-assemblies.
It is well known that the steady-state structure is largely affected by dynamic control, such as the change in temperature and external mechanical forces. 
    For example, Ref.\cite{bupathy_2022} demonstrates the temperature protocol that can select a desired structure from two competing ones in a multicomponent self-assembly.

To design self-assembly structures by dynamic control, we need to access their kinetic pathways, which are unknown from the static interactions.
Systems may often have many metastable states even under the same parameters.
As a result, once the structure gets trapped in the metastable state at a low temperature, the system hardly escapes from it to reach the global minimum.
 Let us take an example of the two-dimensional dodecagonal quasicrystal (DDQC) self-assembled from five-fold symmetrical patchy particles.
 The DDQC can be attained by linearly slowly decreasing temperature in the system (annealing).\cite{lieu_2022b} 
 The obtained structures are not always ideal as the assemblies may have defects.
 This is particularly the case when the speed of temperature change is too fast.
 In this case, the DDQC structure no longer appears.
    In a Monte Carlo simulation of five-patch patchy particles,\cite{vanderlinden_2012} the temperature is quickly cooled down to zero, and then subsequently it is fixed at a specific value. 
    This two-step temperature protocol was developed empirically. 
    The challenge is to find a method that can learn and find suitable temperature settings to facilitate DDQC with few defects, under no or few prior knowledge. 
    In this study, we will show that reinforcement learning is useful for this purpose.
  
Reinforcement learning (RL) is a branch of machine learning that aims to learn an optimal policy and protocol to interact with the environment through experience. 
From the viewpoint of physical science, RL can estimate an external force or parameter change as a function of the state of the system.
Therefore, RL shares many aspects with adaptive optimal control theory.\cite{Bechhoefer:2021} RL can be versatilely applied in strategy games,\cite{silver_2018,openaidota1} robotics,\cite{zhang_2021} and physical problems. 
Applying RL in dynamical physical problems, such as fluid mechanics\cite{Verma:2018,Garnier:2021} and navigation of a single self-propelled particle,\cite{nasiri_2022} is promising because of its capability of finding the best control policy by iterating (experiencing) the dynamical processes without any prior knowledge. 
RL has been applied in optimising the best operational parameters for a system,\cite{huang_2019,zhang_2020} or tuning the operational parameter during a dynamical process.\cite{nasiri_2022} 
In Ref.\cite{zhang_2020}, the Q-learning algorithm\cite{wei_2017} is used to remove grain boundaries from a crystalline cluster of colloids. 
Few studies have focused on many-body particles and their collective behaviours of active matter systems\cite{Norton:2020,Falk:2021,Durve:2020} or self-assemblies.\cite{Whitelam:2020,zhang_2020} 
    In Ref.\cite{Whitelam:2020}, the evolutionary optimisation method has been used to learn temperature and chemical potential changes for self-assembly of complex structures, such as Archimedean tilings. 
    Despite the high performance of this black-box approach, the mechanism of the success remains to be elucidated.
    We will discuss a more detailed comparison between this approach and our method in Sec.~\ref{sec.conclusions}.    
    
    In this study, our main objective is to understand how and why RL works in a self-assembly process. 
    Therefore, we employ a theoretically well-founded algorithm based on Markov decision processes, such as Q-learning, and demonstrate that RL can learn to control the temperature during the self-assembly of patchy particles into DDQC structures. Aside from that, different targets and different models are considered to demonstrate the generality of the proposed RL and to get physical insights for those systems (see Sec.~\ref{sec.conclusions}).
    

    The paper is organised as follows: In Sec.~\ref{sec.methods}, we explain our system and the simulations of the self-assembly, the basics of RL and the Q-learning approach, and the setting of the assembly problem into Q-learning. In Sec.~\ref{sec.results}, we show how the policy is estimated during training, and how the estimated policy works during tests to evaluate its optimality \UL{for the DDQC target}. The generality of the current approach is demonstrated by using different targets whose structures are unknown. 
    In Sec.~\ref{sec.conclusions}, we discuss \UL{several issues, such as how the estimated policy avoids metastable states,} training cost and the discreteness of states in Q-learning. 
    We also discuss physical insights that we get from the RL results and a comparison of different RL approaches.
    Finally, we summarise the main findings of this work.

\section{Methods} \label{sec.methods}
\subsection{Reinforcement learning for dynamic self-assembly}\label{sec.RL.setting}

\begin{figure}[t!]
        \centering
        \includegraphics[width=0.48\textwidth]{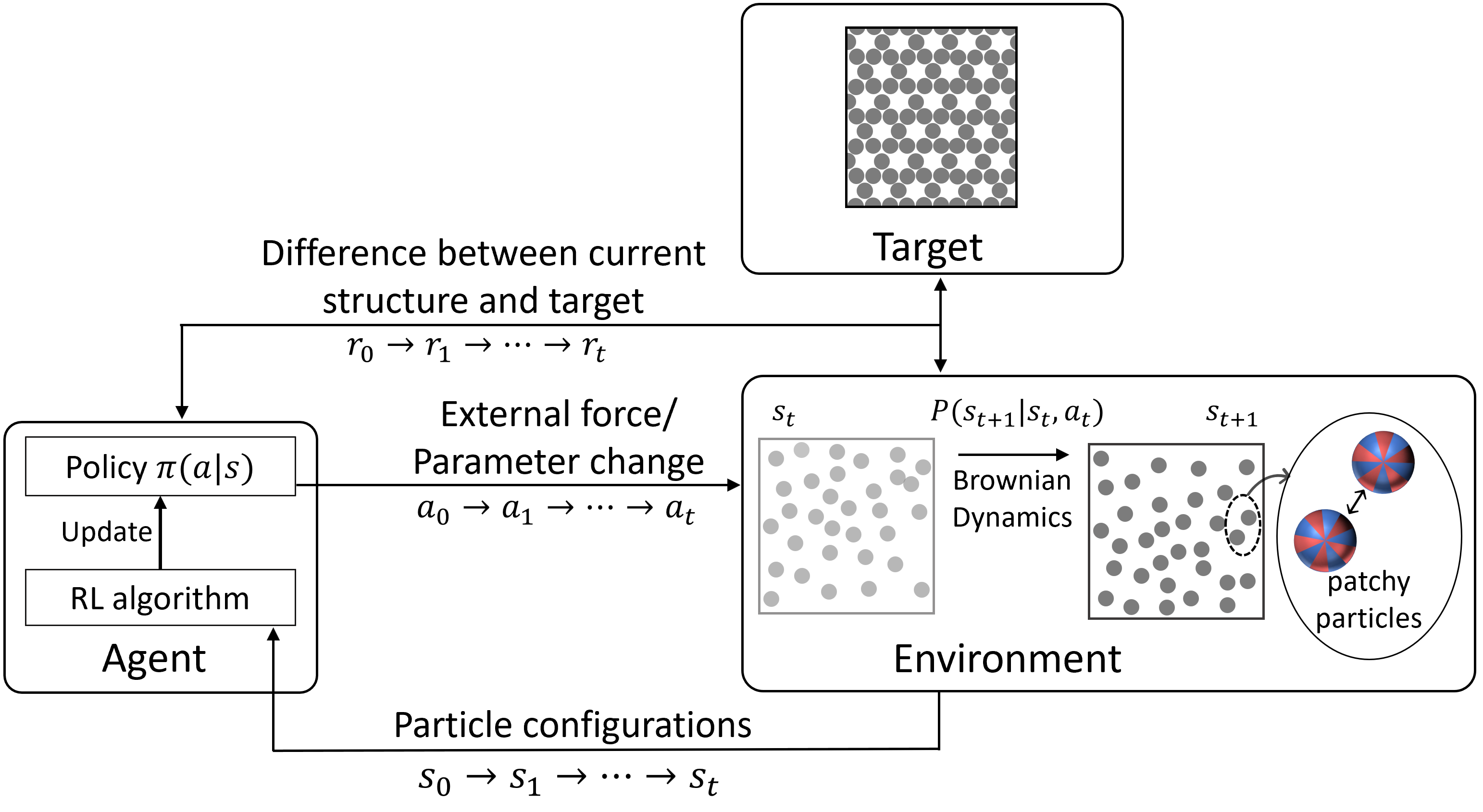}    
        \caption{Schematic of reinforcement learning for dynamic self-assembly. The agent observes the state $s$ from the environment, and decides to take an action $a$ based on the policy $\pi$. The agent learns the policy $\pi$ by a training process to optimise the rewards $r$. In this study, the environment is the particle configuration under a given temperature. The observed states $s$ are the ratio of sigma particle $\sigma$ and the temperature $T$. The action $a$ is to decrease, maintain, or increase the current temperature.}
        \label{fig:RLoverview}
    \end{figure}

    \UL{Figure~\ref{fig:RLoverview} shows the schematic of reinforcement learning (RL) for the dynamical process of self-assembly. 
    In RL, an agent learns how to interact with environments through actions, so as to maximise reward signals.\cite{sutton_1998,brunton_2022,ravichandiran_2020}}
    In the context of self-assemblies, RL aims to control the external force or the parameters \UL{on-the-fly} so that the desired structure is organised from a random particle configuration. 
    In this study, we control the temperature; our action is whether the temperature increases, decreases, or stays at the current value.
    The environment is the configuration of the particles at certain conditions, such as temperature and density. 
    In principle, the dimension of the \UL{particle configuration} is huge.
    It \UL{consists of} all the degrees of freedom of the particles, their positions and orientations\UL{, which are, respectively, $2N$ and $3N$ for the system in this study.} 
    Our purpose is to make the desired structure, which is the DDQC.
    Therefore, we use statistical quantities (or feature values) to characterise the particle configurations.
    This is the number of $\sigma$ particles, denoted as $N_\sigma$; we will discuss this issue in detail in Sec.~\ref{sec.DDQC}.
    We consider two observed states from the environment: the temperature $T$ and the ratio of $\sigma$ particles of the DDQC, which is extracted from the particle configuration, to the total particles.
    We denote the ratio by $\sigma=N_\sigma/N$.
    From the observed states, we take an action $a$ updating the current temperature to the next one. We also get a reward $r_t$ from the measured state. From the reward, the next action is decided at each step and the procedure continues to update all different states. 
    Within each step, the configuration of particles is updated by BD simulations.
    \UL{The control algorithm to be used is Q-learning. The details of RL and Q-learning can be found in Sec.~\ref{secS:RL.and.Qlearning} in Supplementary Information (SI)}.

\begin{figure}[t!]
        \centering
        \includegraphics[width=0.35\textwidth]{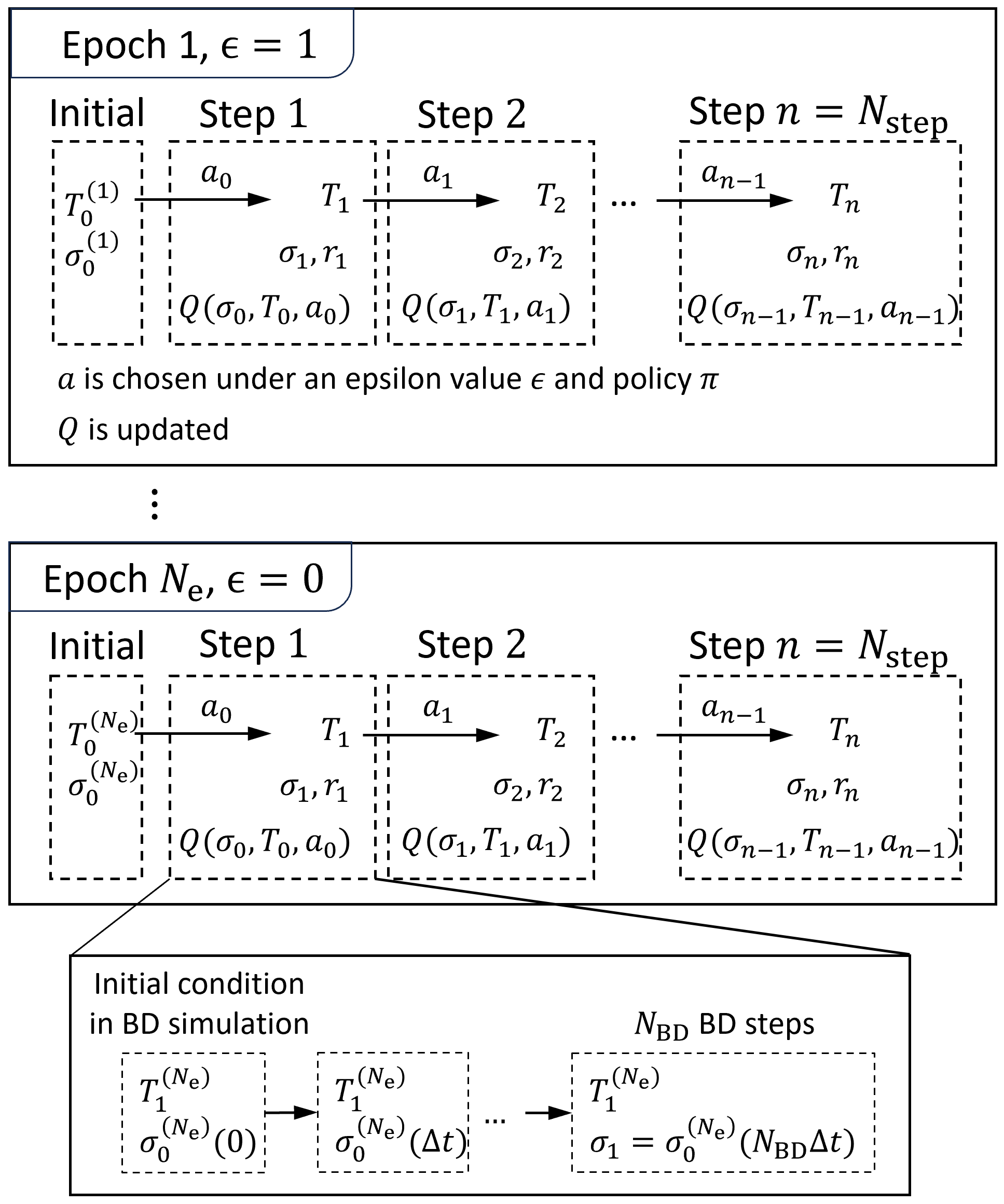}
        \caption{Schematic of Q-learning at each epoch with $\epsilon$-greedy method. The action $a$ is chosen based on the current policy $\pi$ and $\epsilon$. Q is updated according to eq.~\eqref{eq:Qupdate}. Brownian dynamics (BD) simulation is conducted for every action step in $N_\text{step}$ of each epoch.}
        \label{fig:Qlearning}
    \end{figure}

    The schematic for \UL{training with} Q-learning in this study is given in Fig.~\ref{fig:Qlearning}. Initially, Q-table is set to zero for all $a$ and $s$. The RL includes $N_\text{e}$ epochs or episodes in which the $\epsilon$-greedy method is applied. In each epoch, the initial state, i.e. the initial particle configuration and the initial temperature $(\sigma_0,T_0)$ are assigned. Next, the action $a_0$ (either decrease, maintain, or increase $T$) for the temperature is decided based on the current policy and the $\epsilon$-greedy strategy, resulting in the new temperature $T_1$. The Brownian dynamics simulation for the current particle configuration at $T_1$ is conducted. Details of the Brownian dynamics simulation can be found in Sec.~\ref{sec:BD}. The new particle configuration is obtained after a predetermined time $t=N_\text{BD} \Delta t$. Then one can determine the state $\sigma_1$, the reward $r_1$, and eventually update the $Q$-value $Q(\sigma_0,T_0,a_0)$. This concludes the Q-learning of the first step. The next step can be conducted analogically from the current state $(\sigma_1,T_1)$. The Q-table is updated at every action step, every epoch, until the training process ends.

    From the trained Q-table, we can estimate the policy on controlling the temperature with respect to the current state. In order to evaluate the estimated policy, 20 independent tests are conducted. Each test starts with an assigned initial particle configuration and temperature (initial states), followed by consecutive steps of deciding the next action based on the estimated policy, observing the new states, and so on. Otherwise stated, we set the parameters the same as the parameters used during training, except that $\epsilon=0$ is fixed in every test.   

    Table~\ref{table:parameters} shows the parameters of a training set for the target DDQC from patchy particles. The two observed states are the ratio of sigma particle $\sigma$ and the temperature $T$. Initially, the configuration of the particle is random (corresponding to $\sigma_0 \approx 0.1$) and $T_0$ values are chosen randomly in the investigated range. \UL{The initial positions are obtained by inflation of randomly distributed points under repulsive force to prevent overlapping. During RL, w}hile the fraction of sigma never reaches out of the range $[0,1]$, the temperature $T_{k+1}$ after the action $a_{k}$ may exceed the investigated range. In this case, the updating is carried out as usual except that we treat $T_{k+1}=T_{k}$. The policy after training is used for the test at the same conditions as training (except $\epsilon$).

    \begin{table}[h]
    \small
    \centering
    \setlength{\tabcolsep}{4pt}
    \setlength\extrarowheight{0pt} 
    \begin{tabularx}{0.48\textwidth}{|L|L|} 
        \hline
        \textbf{Parameter}   &  \textbf{Value}  \\
        \hline
        - States of sigma fraction, $\sigma$  &   $[0,1]$ with an interval of 0.1\\
        - States of temperature, $T$          &   $[0.2,1.3]$ with an interval of 0.1\\
        - Actions on the temperature, $a$      &   $\{-0.05, 0, 0.05\}$\\
        - Number of epochs, $N_\text{e}$      & 101  \\
        - $\epsilon$-greedy                   & Linearly decrease in each epoch from 1 to 0 \\
                                            
        - Initial temperature at each epoch, $T_0$    & Random, $T_0 \in [0.2, 1.3]$ \\
        - Initial structure at each epoch, $\sigma_0$  & Random configuration  ($\sigma_0\approx0.1$) \\
        -Number of action steps \UL(RL steps) in each epoch, $N_\text{step}$ & 200\\
        - Number of BD steps in each  \UL{RL step}, $N_\text{BD}$ & $100,000$ steps \\
        \UL{- Time step in BD simulation, $\Delta t$ }    & \UL{$10^{-4}$}  \\
        - Target, $\sigma^*$  &   0.91 \\
        - Rewards, $r$  &   $-(\sigma-\sigma^*)^2$      \\   
        - Learning rate, $\alpha$ & 0.7\\
        - Discount factor, $\gamma$ & 0.9\\        
        - Number of particles, $N$ & 256\\
        - Area fraction, \UL{$A=\pi a^2 N/(L_x L_y)$}   &   0.75 \\        
        \hline                        
    \end{tabularx}
    \caption{Parameters for the training set of DDQC patchy particles.} 
    \label{table:parameters}
\end{table}

\subsection{Self-assembly of patchy particles through Brownian Dynamics simulations} \label{sec:BD}

    Our system consists of $N$ patchy particles. Each particle stochastically moves and rotates following the equations \eqref{BD.x} and \eqref{BD.angle} under the temperature $T$ at time $t$ (see Fig.~\ref{fig:RLoverview}).
    The patchy particle has anisotropic interactions with other particles.
    Depending on the anisotropy, the particles may form an ordered self-assembled structure.
    Because the thermal fluctuation of the particles is dependent on the temperature, the self-assembled structure varies as the temperature changes.    

    The Brownian dynamics (BD) simulation is employed to simulate the assembly of five-fold-symmetric patchy particles.\cite{lieu_2022a,lieu_2022b} The patchiness on the spherical particle is described by the spherical harmonic of $Y_{55}$. There are 5 positive patches and 5 negative patches arranged alternatively around the particle's equator (see Fig.~\ref{fig:RLoverview}). We set that the same sign patches are attractive while opposite patches are repulsive. The particles are confined to a flat plane, meaning that the particles can translate on the plane while they can rotate freely in three dimensions. \UL{We use $NVT$ ensemble with periodic boundary condition applied on the $x$ and $y$ direction of a simulation box of the size $L_x \times L_y \times 2a$, where $a$ is the particle radius.} In the Brownian dynamics, the position $\mathbf{r}$ and orientation $\bm{\Omega}$ of the particle are updated according to the equations
    \begin{equation}
	 \mathbf{r}(t+\Delta t)
	 = \mathbf{r}(t) 
	 + \frac{D^\text{T}}{k_\text{B} T} \mathbf{F}(t) \Delta t 
	 + \delta \mathbf{r} , 
    \label{BD.x}
    \end{equation}
    
    \begin{equation}
	 \bm{\Omega}(t+\Delta t)
	 = \bm{\Omega}(t) 
	 + \frac{D^\text{R}}{k_\text{B} T}  \mathbf{T}(t) \Delta t 
	 + \delta \bm{\Omega},
    \label{BD.angle}
    \end{equation}
    where $D^\text{T}$, $D^\text{R}$ are the translational and rotational diffusion coefficients, respectively, $k_\text{B}$ is the Boltzmann constant, $\mathbf{F}$ and $\mathbf{T}$ are the force and torque, the Gaussian noise terms $\delta \mathbf{r}$ and $\delta \bm{\Omega}$ are with zero mean and satisfying 
    $\langle \delta \mathbf{r} \delta \mathbf{r}^\intercal\rangle=2 D^\text{T} \Delta t$ and 
    $\langle \delta \bm{\Omega} \delta \bm{\Omega}^\intercal\rangle=2 D^\text{R} \Delta t$, respectively.
    The characteristic length, energy, time, and temperature for the nondimensionalisation are the particle radius $a$, the potential well-depth $\varepsilon_\text{V}$, the Brownian diffusion time $\tau_\text{B}=a^2/D^\text{T}$, and $\varepsilon_V/k_B$, respectively.   

    The interaction potential of a pair of particles $i$ and $j$ is $V_{ij}=V_{\text{WCA}}(r_{ij}) + V_{\text{M}}(r_{ij})\Xi(\bm{\Omega}_{ij})$. The isotropic Week-Chandler-Anderson term $V_{\text{WCA}}$ prevents the overlapping of particles. The interaction of the patchiness is given by the Morse potential $V_{\text{M}}$ and the mutual orientation dependent term $\Xi(\bm{\Omega}_{ij})$. 
    \begin{equation}
	V_{\text{WCA}}= 
  	\begin{cases}
  	4\varepsilon_\text{V} \left[ (\frac{2a}{r})^{12}- (\frac{2a}{r})^{6}+\frac{1}{4} \right]	, & r\leq 2a\sqrt[6]{2} \\
  	0 												    							, & r > 2a\sqrt[6]{2}
  	\end{cases} 
    \end{equation}

    \begin{equation}
	V_\text{M}= \varepsilon_\text{V} M_\text{d} \left \{ \left[ 1-e^{\left( -\frac{r-r_{\text{eq}}}{M_\text{r}} \right)} \right]^2 -1 \right \} ,
    \end{equation}
    where $r$ is the center-particle distance; is the Morse potential equilibrium position, depth and range are respectively $r_{\text{eq}}=1.878a$, $M_\text{d}=2.294a$, and $M_\text{r}=a$.\cite{delacruz-araujo_2016}

    The orientation of particle $i$ is determined by the orthogonal local bases $\nvec^{(i)}_m$, $m=1,2,3$. Let $\rvec$ be the unit distance vector of particle $i$ and $j$. The interaction of a pair of particle $Y_{lm}$ is $\Xi_{lm}  \propto \{ \nvec_0^{l-m} \nvec_+^{m}\}_{(i)} \odot   \{\rvec^{2l} \}   \odot  \{ \nvec_0^{l-m} \nvec_+^{m}\}_{(j)} $, where $\nvec_0=\nvec_3$, $\nvec_+=\frac{1}{\sqrt{2}}(\nvec_1 + i \nvec_2)$, and the $\{\}$ indicates the irreducible tensor. For a pair of $Y_{55}$ particles $\Xi_{55}  \propto \{  \nvec_+^{5}\}_{(i)} \odot   \{\rvec^{10} \}   \odot  \{ \nvec_+^{5}\}_{(j)} $, and $\Xi$ is normalised to be in the range of [-1,1].

\subsection{Characterisation of DDQC structures}\label{sec.DDQC}
    
\begin{figure}[t!]
        \centering
        \includegraphics[width=0.48\textwidth]{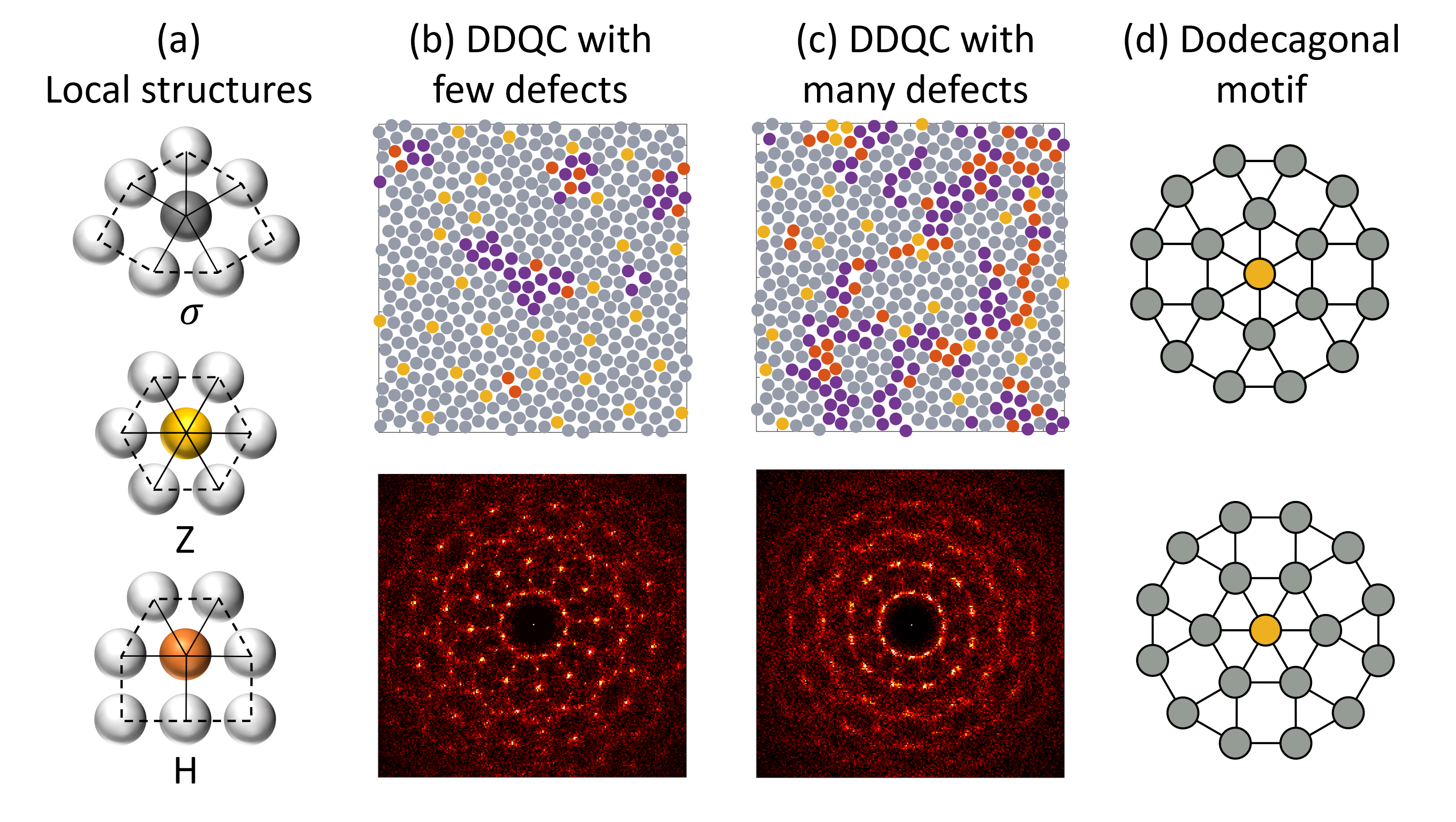}
        \caption{Characterisation of DDQCs. (a) Demonstration of local structures. (b-c) Examples of DDQC with few and many defects, and the correspondent Fourier transformations. The \UL{undefined particles ($U$)} are marked purple. The fraction of sigma in (b-c) are 0.84 and 0.67, respectively. (d) Dodecagonal motif made from one $Z$ particle centred in 18 $\sigma$ particles.}
        \label{fig:DDQC}
    \end{figure}

    One method to characterise two-dimensional DDQC is to determine local \NY{structures} around each particle according to its nearest neighbours \cite{vanderlinden_2012,reinhardt_2013,lieu_2022b} (Fig.~\ref{fig:DDQC}). Given the particle positions, the $\sigma$, hexagonal $Z$, and $H$ local structures are estimated. A DDQC structure usually contains a few $Z$ dispersed in many $\sigma$ and a few $H$ particles. In detail, dodecagonal motif, which is made from one centred $Z$ and 18 $\sigma$ particles (Fig.~\ref{fig:DDQC}d), is observed in the DDQC. The motifs can be packed in different ways, e.g. the centres form triangles\UL{ and squares.\cite{vanderlinden_2012,lieu_2022b}} The ratio of the $\sigma$, $Z$, and $H$ particles to the total particles in the packed motifs are found to be $0.8 \leq \sigma \leq 0.93$, $0.07 \leq Z \leq 0.14$, and $0 \leq H \leq 0.13$, respectively. 
    Such ratios are found comparable to those in square-triangle tiling\cite{leung_1989} or simulated DDQC.\cite{vanderlinden_2012,lieu_2022b} 

    In our previous study,\cite{lieu_2022b} we found that DDQC consists of several different motifs, each of which can form an approximate.
    In this view, the ratio of \NY{the number of} sigma particles to the total \NY{number of particles} in DDQCs is expected to be $0.8 \leq \sigma \leq 0.93$. However, at finite temperatures, defects can always appear during the self-assembled process, and therefore, $\sigma$ can be smaller than those values.
    Besides that, when the structure is frozen with defects and forms metastable states, $\sigma$ becomes much smaller.  

        \UL{
        In order to clarify equilibrium and metastable structures, RL is supplemented by three additional simulation methods
        under the same conditions of RL with area fraction $A=\pi a^2N/(L_x L_y)= 0.75$, and the investigated temperature $T\in[0.2, 1.3]$. 
        The three methods are (i) Replica Exchange Monte Carlo\cite{sugita_2000,iba_2001} (REMC, or called parallel tempering) described in Sec.~\ref{secS:REMC} in SI, (ii) BD simulations under quenching to a fixed temperature (Sec.~\ref{secS:quenching} in SI), and (iii) BD simulations under annealing temperature (Fig.~\ref{fig:RL.vs.annealing}).
        In (ii), we prepare an initial configuration at random position and orientation, and then, set the temperature at a lower value.
        }

    \UL{
     In REMC, we find that the equilibrium phases are fluid at high temperature $T \gtrsim 1.8$, the $Z$-phase at intermediate temperature $T \gtrsim 0.9$, DDQC at low temperature $T \lesssim 0.89$.    
    We should note that the $\sigma$-phase appears when the area fraction is smaller than the current value.
    Details of the calculation of the phase diagram can be found in Fig.~\ref{fig:REMC} and Sec.~\ref{secS:REMC}.
    }  
        
    \UL{
    The DDQC with defects can typically be seen in the self-assemblies under quenching to a fixed temperature from a random initial configuration (Fig.~\ref{fig:S.quenching}). 
    At intermediate $T\in [0.7,0.85)$, there is DDQC with $\sigma \in [0.7,0.9]$. At lower temperature $T\in [0.35,0.7)$, $\sigma$ decreases within $\sigma \in [0.5,0.7]$ as $T$ decreases. 
    This lower value of $\sigma$ is different from that of the equilibrium state, as shown in Fig.~\ref{fig:REMC} (see also Table~\ref{table:S.compare.structure}).
    Therefore, we refer to these structures as metastable because many particles are kinetically trapped during the growth of DDQC. At much lower T, $\sigma$ keeps decreasing to 0.2.
    } 
    
    \UL{During the training of RL, the value of $\sigma$ 
    is used to evaluate the DDQC structures (Fig.~\ref{fig:DDQC}).} 
    We consider that structures with $0.7 \lesssim \sigma$ \UL{$ \lesssim 0.93$} are global free energy minimum DDQCs with a few defects, \UL{because those values of $\sigma$ are comparable with the ideal value of DDQC and also the value of equilibrium DDQC (Table~\ref{table:S.compare.structure}).} 
    On the other hand, when $0.5 \leq \sigma \leq 0.7$, there are many defects in the structures and we refer to them as metastable states. 
    \NY{This argument confirms that the global free energy} minimum DDQC and metastable structures can be distinguished by the value of $\sigma$. 
    \NY{In fact, their} Fourier transformation is distinguishable.
    The DDQC has clear 12-fold symmetric spots separated from the background, whereas the metastable structure has blurred 12-fold symmetry (Fig.~\ref{fig:DDQC}).

\subsection{Target structures for RL}\label{method.targetRL}

   \UL{In RL,} we choose $\sigma$ as one of the RL states with $0\leq \sigma \leq 1$. The value $\sigma^*$ of the target DDQC is set as $\sigma^*=0.91$. 
    \UL{At the density used in this study, using $\sigma$ alone is adequate to identify different states formed during the assembly process (see Table~\ref{table:S.compare.structure} for the details of other local structures). In other cases where the system has more complex structures, }
    other quantities, such as $Z$ and $H$, may be needed. 
    However, as we demonstrate, using $\sigma$ \UL{not only works for DDQC target $\sigma^*=0.91$ (Sec.~\ref{RL.DDQC}), but also for other targets such as $\sigma^*=0.65$ and $\sigma^*=0.35$ (Sec.~\ref{RL.othertarget})}.
    \NY{We will demonstrate that RL is capable of stabilising a metastable structure ($\sigma^*=0.65$) and even finding a policy to control the structure dynamically to realise the unstable target structure ($\sigma^*=0.35$).
    To do this, we use the value iteration method for the two targets (see Sec.~\ref{secS:value.iteration} in SI).
    }
    
    Another state used in RL is the temperature $T$. The range of the temperature is chosen as $0.2\leq T \leq 1.3$ so that the particle interaction dominates the noise at $T_\text{min}$ and the noise dominates the interaction at $T_\text{max}$. 

\NY{
To clarify how RL overcomes the energy barriers to reach the global minimum, we also study the model of \UL{a point particle in} a triple-well potential in Sec.~\ref{sec.conclusions} and Sec.~\ref{secS:triple.well}. 
We also demonstrate the generality of proposed RL by estimating the policy for DDQC formation from particles interacting through a two-lengthscale isotropic potential in Sec.~\ref{sec.conclusions} and Sec.~\ref{secS:iso}.}
It is known that this model also exhibits DDQC.\cite{engel_2007}
Details of the isotropic potential can be found in Ref.\cite{lieu_2022b} and the references therein.


\section{Results} \label{sec.results}

\subsection{Optimal temperature change to generate DDQC from patchy particles}\label{RL.DDQC}
\subsubsection{\UL{Training process}}

    \begin{figure*}[!]
        \centering
        \includegraphics[width=0.9\textwidth]{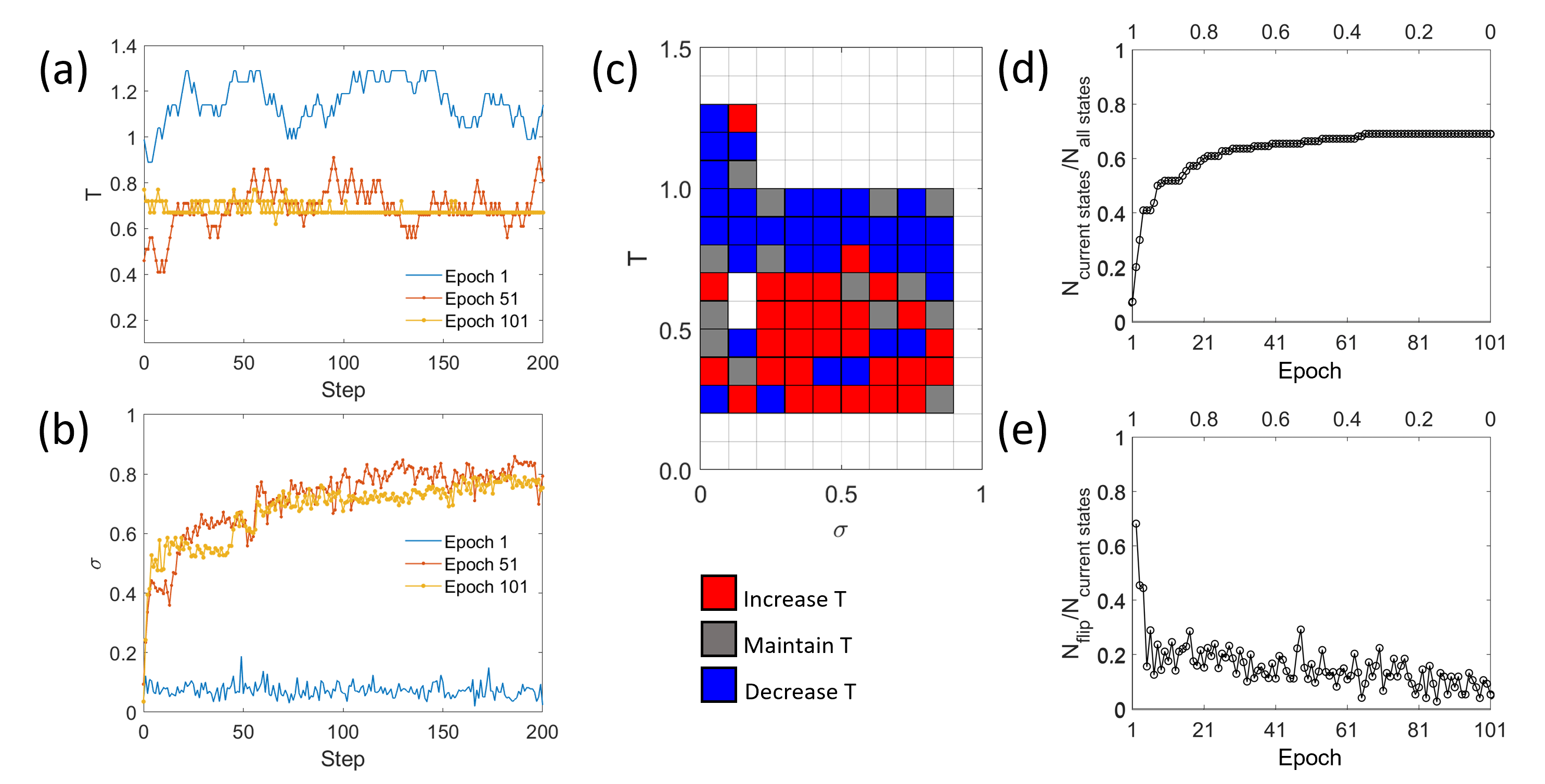}
        \caption{Training data at the condition of random $T_0$ and number of epochs $N_\text{e}=101$ in Table~\ref{table:parameters}. (a,b) The progression of the states $T$ and $\sigma$ at selected epochs: first, middle and last epoch (equivalent $\epsilon=0, 0.5, 1$ respectively); (c) the policy after training; (d) the change of ratio of the number of accessed states to total states and (e) ratio of flipped-policy states to accessed states after each epoch during training, the horizontal axis on the top of the graph is the corresponding value of $\epsilon$.}
        \label{fig:allT0Ne101.training}
    \end{figure*}

        \begin{figure*}[!]
        \centering
        \includegraphics[width=1\textwidth]{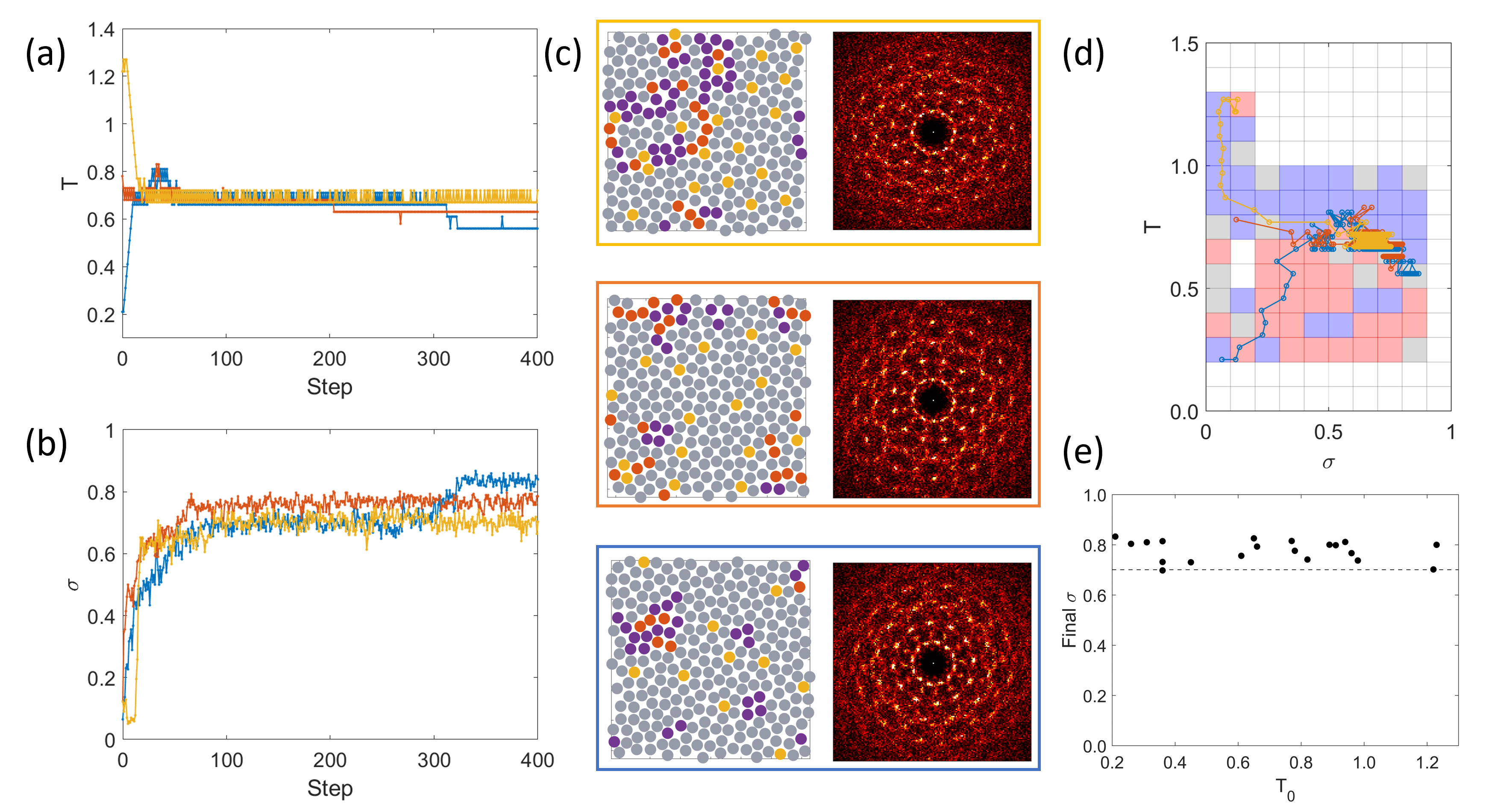}
        \caption{Testing data of the policy obtained at the condition of random $T_0$ and number of epochs $N_\text{e}=101$ in Fig.~\ref{fig:allT0Ne101.training}. Samples starting with \UL{low (blue), intermediate (red), high (yellow) initial temperature} are shown with (a) the temperature schedule, (b) corresponding $\sigma$, and (c) snapshots at the last step of the corresponding trajectories. (d) The trajectories of (a,b) on the policy plane obtained from Fig.~\ref{fig:allT0Ne101.training}(c), in which the starting points are from the left side. Changes of temperatures of the trajectories follow the policy shown in the background. (e) The dependence of $\sigma$ on the initial temperature $T_0$ obtained from 20 independent samples. The dashed line is a guide to the eye for the lower limit of global minimum DDQCs.}
        \label{fig:allT0Ne101.testing}
    \end{figure*}

    First, we demonstrate the capability of Q-learning to find the best temperature schedule to create DDQCs of patchy particles from random configurations. Figure~\ref{fig:allT0Ne101.training} shows the training result under the condition in Table~\ref{table:parameters}, where the policy is trained with $N_\text{e}=101$ epochs and the initial temperature $T_0$ at each epoch is randomly assigned within the investigated range. 
    During each epoch, the action changes according to the current policy and $\epsilon$, hence the states of temperature $T$ and ratio $\sigma$ change at each step, as shown in Fig.~\ref{fig:allT0Ne101.training}(a-b). 
    In the first epoch in which $\epsilon=1$, $T$ fluctuate around $T \approx 1.0$. 
    Accordingly, $\sigma$ is low $\sigma<0.2$ and far from the target value $\sigma^*$. As the training continues, Q-table is updated. 
    \UL{At the next epoch, we repeat the process with a new initial random configuration and random temperature. However, we use smaller $\epsilon$ and updated Q-table.}
    In the mid epoch $n=51$ at which $\epsilon=0.5$, $T$ fluctuates around $T \approx 0.7$ whereas in the last epoch $n=101$ at which $\epsilon=0$, $T$ shows less fluctuation around $T \lesssim 0.7$.
    After the epoch $n=51$, $\sigma$ approaches closer to $\sigma^*$.

    Figure~\ref{fig:allT0Ne101.training}(c) demonstrates the policy after training, which is the action for the maximum of $Q$, namely, $\argmax_{a} Q(\sigma,T,a)$. This state-space roughly consists of two regions divided by a \UL{\textit{characteristic temperature}} $T^*=0.7$. 
    The estimated policy is to decrease the temperature above $T^*$, and to increase the temperature below $T^*$. 
    When $\sigma \geq 0.8$, the temperature can be decreased further to $T\approx 0.5$. 
    The action of `maintaining temperature' can be seen in the policy, but no clear correlation to the states is observed. The policy has states that are not accessed during training. The action for these inaccessible states is random. Figure~\ref{fig:allT0Ne101.training}(d) presents the ratio of the number of accessed states to total states during training (total number of states is $10\times11$). 
    We also measure whether the policy converges to its optimal in Fig.~\ref{fig:allT0Ne101.training}(e), by defining the ratio of the number of flipped states to accessed states. The flipped state is counted when the policy at the current epoch $\pi^{(i)}(s,a)$ changes compared to the policy at the previous epoch $\pi^{(i-1)}(s,a)$. 
    The ratio decays to $\lesssim 0.1$, but the decay is slow.
    Even after the epoch of $\epsilon \lesssim0.4$ after which the number of accessed states reaches a plateau, the ratio is still decaying slowly.
    This result suggests that many epochs are required to reach an optimal policy.

\subsubsection{\UL{Testing evaluation}}

    After training, the estimated policy is tested. 
    The results of the test are presented in Fig.~\ref{fig:allT0Ne101.testing}. 
    The time evolution of temperature and $\sigma$ during the test with initial configurations of random particle positions and orientations and with random $T_0$ are shown in Fig.~\ref{fig:allT0Ne101.testing}(a,b). 
    At first, $T$ quickly reaches \UL{the characteristic temperature $T^*=0.7$}, then fluctuates around that value until $\sigma$ reaches the target value. 
    Finally, $T$ decreases at a considerably slower rate to $T\approx 0.6$.
    The final temperature is dependent on each realisation; in some cases, $T$ reaches $T\approx 0.6$, whereas, in other cases, $T$ stays at $T\approx 0.7$. 
    Correspondingly, the final value of $\sigma$ is either $\sigma \approx 0.8$ or slightly smaller than that.
    The snapshots at the final steps have dodecagonal motifs consisting of one $Z$ particle centred in 18 $\sigma$ particles (see Fig.~\ref{fig:DDQC}(d)).
    The intensities in the Fourier space show clear twelve-fold symmetry, although some defects are present in the real space. 
    
    Figure~\ref{fig:allT0Ne101.testing}(d) shows trajectories of the states ($T$ and $\sigma$) during the test together with the estimated policy.
    We show the three trajectories with different initial temperatures: high $T_0$, intermediate $T_0$, and low $T_0$.
    In the case of high $T_0$, the temperature decreases to $T \approx 0.8$ but $\sigma$ does not increase.
    Once the temperature becomes $T \approx 0.7$, dodecagonal structures start to appear and $\sigma$ increases and fluctuates around $\sigma\approx 0.7$.  
    
    In the case of low $T_0$, some dodecagonal structures appear from the beginning because the temperature is low.  
    As the temperature is increased to $T \approx 0.7$, $\sigma$ is also increased and reaches $\sigma \approx 0.7$.  
    The temperature is found to decrease at the point $\sigma \approx 0.8$.  
    
    When the initial temperature $T_0$ is intermediate, $\sigma$ increased, then fluctuates, and finally, it is increased more when $T$ is decreased slightly. 
    Note that in all cases, the initial $\sigma$ is small because the initial configuration of particles is random in position. 
    Using this policy, the DDQC structure can be obtained in tests at any value of the initial temperature $T_0$ (Fig.~\ref{fig:allT0Ne101.testing}(e)). 
    In short, the RL agent has found out the role of the \UL{characteristic} temperature $T^*$ in facilitating the formation of the DDQC structure. As a result, when the DDQC is not formed (low $\sigma$), the RL policy suggests to drive the temperature to $T^*$ until a DDQC (high $\sigma$) is formed, then decrease $T$ to stabilise the structure. 
    \NY{
    We discuss $T^*$ in comparison with the phase transition temperature at equilibrium states in Sec.~\ref{sec.REMC}.
    }
        The RL has discovered the \UL{characteristic} temperature by itself. We do not feed any information about the role of $T^*$ or its value. 

\begin{figure}[!]
        \centering
        \includegraphics[width=0.48\textwidth]{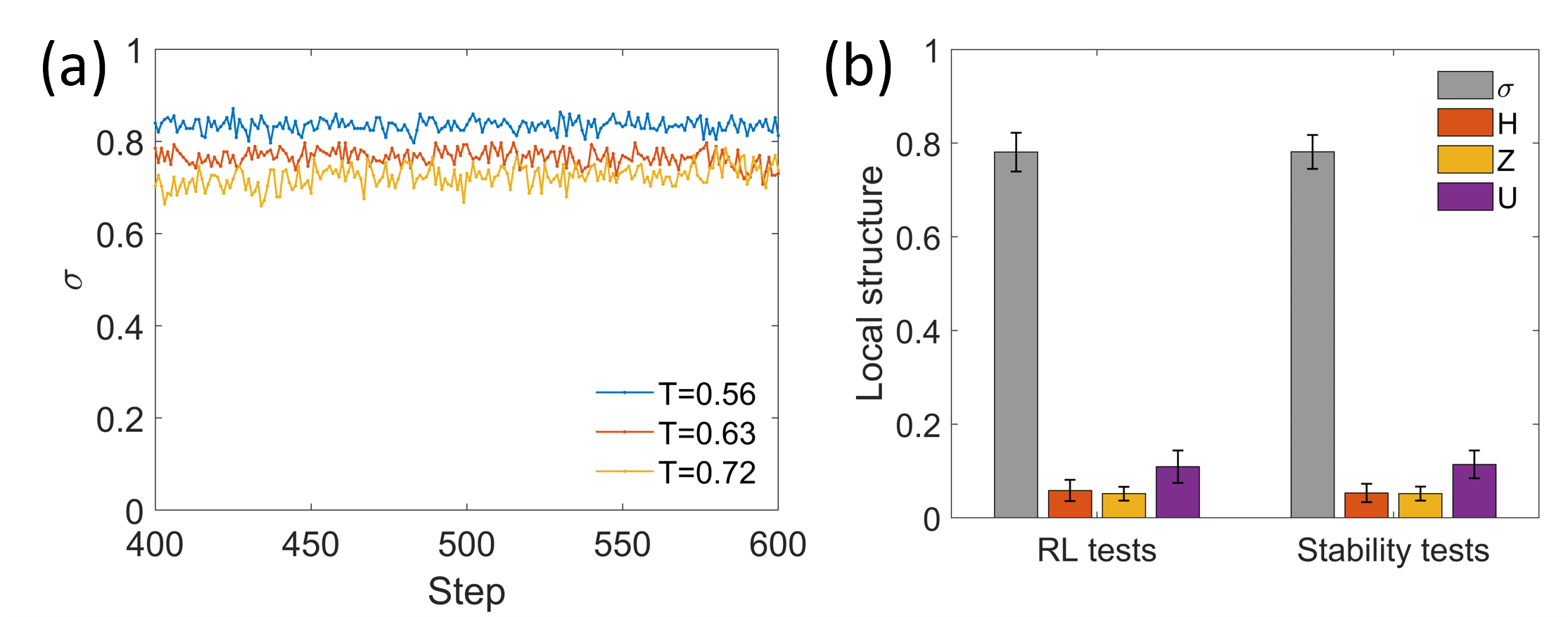}
        \caption{\UL{Stability of the RL tests at the condition of random $T_0$ and number of epochs $N_\text{e}=101$ in Fig.~\ref{fig:allT0Ne101.testing}. (a) Continuation of the optimised structures from the step 400 in Fig.~\ref{fig:allT0Ne101.testing}(a) at fixed temperature. The number of BD steps is $20\times10^6$ steps, equivalent to 200 RL steps. (b) Statistics of the local structure for 20 independent samples in Fig.~\ref{fig:allT0Ne101.testing}(e).}}
        \label{fig:stabilise.DDQC}
    \end{figure}

    \UL{
    We further check the stability of the optimised structures, by extending the simulations under the fixed temperatures at the last RL step.
    The results of the stability test are shown in Fig.~\ref{fig:stabilise.DDQC} along with the statistics of the local environments.
    Compared with the RL test, we found no significant difference in the statistics of the local environments, which implies that optimised structures are inherently stable.   
    }

\subsubsection{\UL{Comparison of RL with conventional approaches}}
\begin{figure}[t!]
        \centering
        \includegraphics[width=0.3\textwidth]{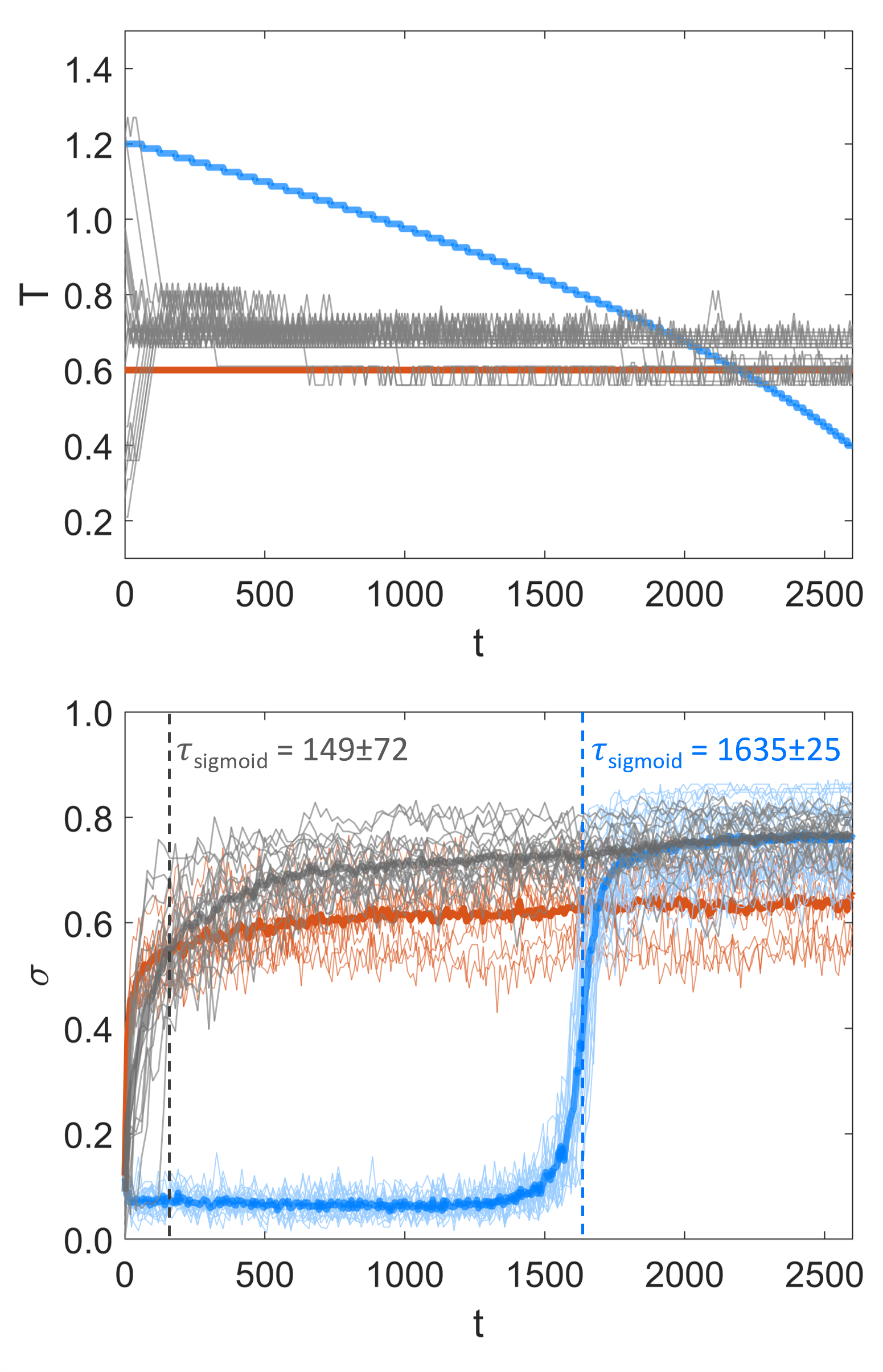}       
        \caption{Comparing DDQC assemblies by RL temperature policy and annealing. The thin lines are the temporal changes of temperature and $\sigma$ ratio of samples from RL testing (grey), annealing (blue), \UL{and $T=0.6$ (red)}. The bold lines of $\sigma$ are the mean values. Simulation parameters are $N=256$, $A=0.75$, random initial configuration. \UL{The dashed line $\tau_\text{sigmoid}$ can be considered as the average onset of DDQC for RL (grey) and annealing (blue) tests fitted by sigmoid functions.}}
        \label{fig:RL.vs.annealing}
    \end{figure}

    Next, we compare the formation of a DDQC using the estimated policy with the self-assembly using the conventional annealing \NY{and quenching methods}.\cite{lieu_2022b} 
    Figure~\ref{fig:RL.vs.annealing} shows the trajectories of $T$ and $\sigma$ for different realisations.
    In the annealing simulations, we have used the linear temperature decrease during BD steps.
    In this case, the time step for each BD step was also decreased for numerical stability.
    In annealing and RL methods, $\sigma$ values reach $\sigma \approx 0.8$, at which the dodecagonal structures appear clearly with a few defects. \UL{The ratios of $Z$ and $H$ are also comparable between the two methods (see Table~\ref{table:S.compare.structure}).} \NY{To evaluate the speed of DDQC formation, we fit each trajectory of $\sigma(t)$ by sigmoid function and estimate its time.}
    In the case of the annealing, we have used a pre-fixed temperature schedule, and therefore, $t \gtrsim 1600$ is required for the dodecagonal structures. 
    \NY{Before the temperature reaches $\approx 0.8$, no structural changes occur.}   
    \UL{In contrast, with the RL policy, $\sigma$ increases quickly and then levels off. The estimated timescale is $t\approx 150$, which is much faster than that of annealing.
    We also compare RL and rapid quenching at the fixed temperature. 
    The temperatures in RL change, but their values finally become in the range of $[0.6, 0.7]$. Therefore, the quenching temperature we chose is $T=0.6$. As depicted in Fig.~\ref{fig:RL.vs.annealing}, the obtained assemblies have $\sigma \approx 0.6$, meaning that they are trapped at the metastable states and contain more defects than that of RL or annealing. Readers may refer to Fig.~\ref{fig:S.quenching} for the quenching at other temperatures.
    } 

\begin{figure}[!]
        \centering        
        \includegraphics[width=0.48\textwidth]{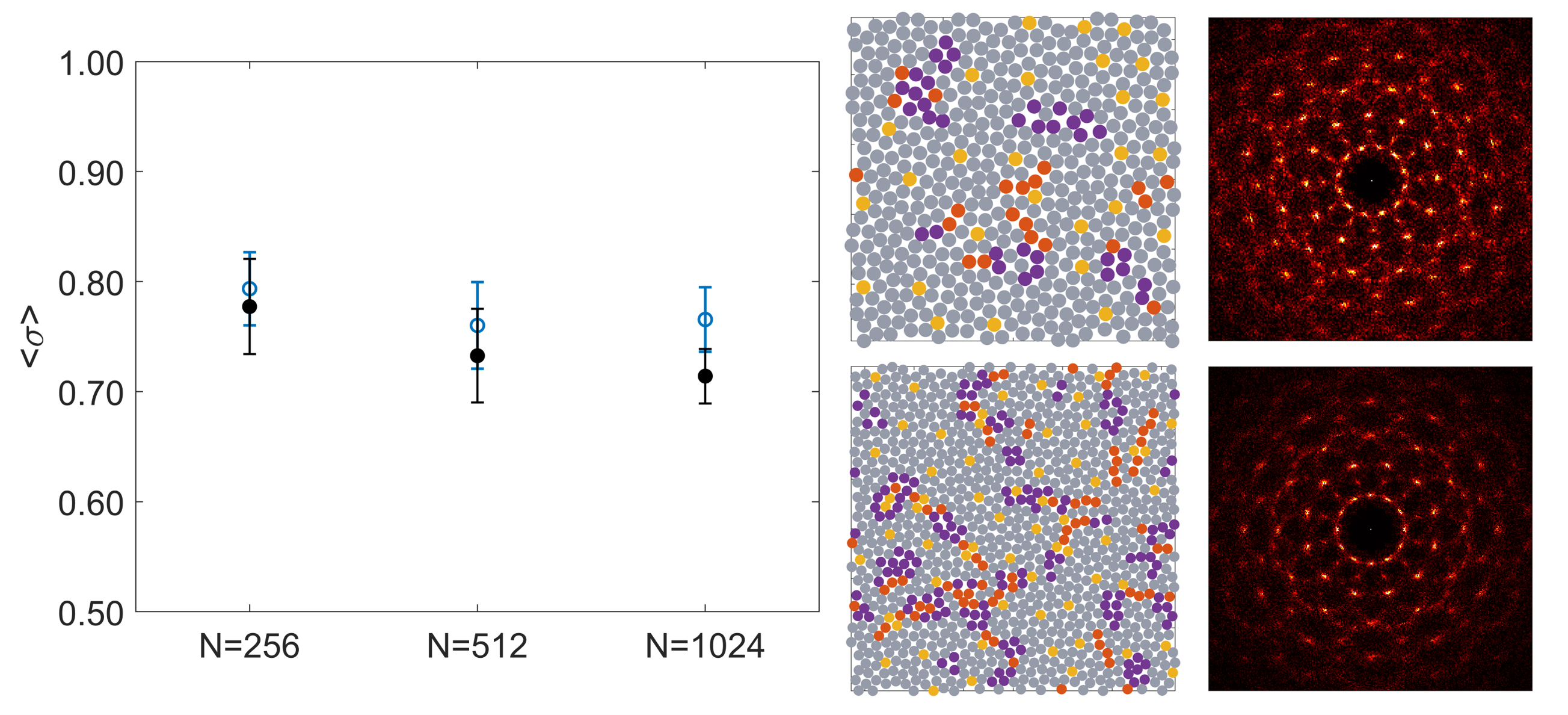}
        \caption{Performance of the testing on different system sizes, where the policy is trained with the system of $N=256$ particles. The means and standard deviations are calculated from 20 independent samples \UL{after 400 RL steps (black) 1000 RL steps (blue)}. The snapshots and Fourier transformations are demonstrated for $N=512$ (upper) and $N=1024$ (lower) tests.}
        \label{fig:test.allT0.systemsize}
    \end{figure}

    In the training steps, we use a small system size, $N=256$. 
    It is important to check whether the estimated policy using RL can work upscale. 
    We perform tests at larger system sizes $N=512$ and $N=1024$. 
    Figure~\ref{fig:test.allT0.systemsize} demonstrates the obtained structures of different system sizes. 
    The estimated policy for the smaller systems size works even for the tests with all investigated system sizes, namely, we obtain $\sigma \gtrsim 0.7$. The mean value of $\sigma$ seems to slightly decrease with system size. This is because the larger system size requires more time to stabilise. If more steps are conducted for a larger system size, there is no significant difference difference the three groups. 
    In fact, the snapshots both in the real and Fourier spaces for the larger system sizes show dodecagonal structures. 

\subsection{Reinforcement learning for unknown targets of patchy particles}\label{RL.othertarget}
    
    \begin{figure*}[!]
        \centering
        \includegraphics[width=1\textwidth]{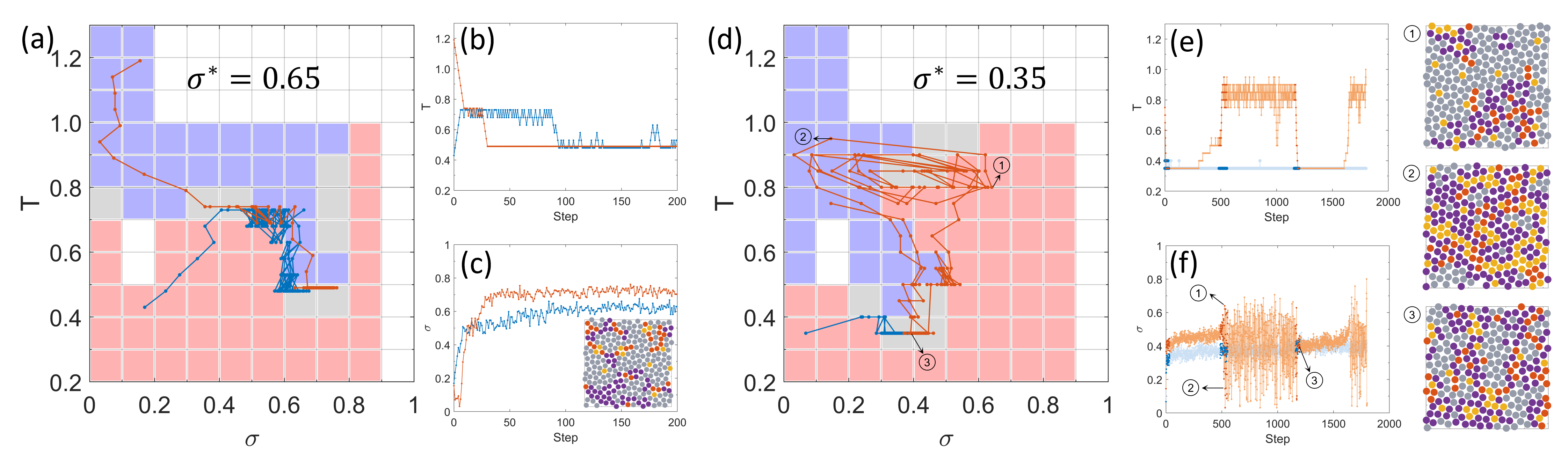}
        \caption{Reinforcement learning for other assemblies of patchy particles with the targets \UL{(a-c)} $\sigma^*=0.65$ and \UL{(d-f)} $\sigma^*=0.35$. (a-c) The policy with selected trajectories during the tests and the corresponding temperature and $\sigma$ of the tests for $\sigma^*=0.65$. The mean and standard deviation of $\sigma$ from 20 independent samples are 0.63 and 0.03, respectively. The snapshot at the last point of the blue trajectory is given. (e-f) The temperature schedule and $\sigma$ of selected trajectories for $\sigma^*=0.35$. Three snapshots of the orange trajectory are given. (d) The trajectories of the tests on the policy plane in which only part of the data (darker points) is used. The mean and standard deviation of $\sigma$ from 20 independent samples are 0.42 and 0.14.}
        \label{fig:patchy.othertarget}
    \end{figure*}
    
    In \UL{Sec.~\ref{RL.DDQC}}, we use the target $\sigma^*=0.91$ to obtain the DDQC structure.
    This structure \NY{is at an equilibrium state} under a certain temperature \NY{(see Sec.~\ref{sec.REMC})}.
    In this section, we demonstrate that RL also works for the unknown target structures, which are not equilibrium states.
    To do this, we perform RL for different targets: $\sigma^*=0.65$ and $\sigma^*=0.35$ in patchy particle systems. The estimation of the policy is conducted by the value iteration method (Sec.~\ref{secS:value.iteration}) instead of training the Q-table through numerous episodes because of the availability of sufficient data (see Table~\ref{table:train.DDQC.batch}). 
    As shown in Fig.~\ref{fig:patchy.othertarget}, RL estimates different policies for different targets.
    
    For $\sigma^*=0.65$, the structure obtained from the estimated policy \UL{has $\sigma\approx \sigma^*$, which} is close to DDQC but with many defects. 
    The policy in Fig.~\ref{fig:patchy.othertarget}(a) shows a border at the \UL{characteristic} temperature at $T\UL{^*}=0.7$ when $\sigma$ is small $\sigma < 0.6$. The policy is similar to the case of \UL{DDQC target} in Fig.~\ref{fig:allT0Ne101.training}-\ref{fig:allT0Ne101.testing}. It suggests to drive the temperature to $T^*$ so that $\sigma$ increases, that is to decrease $T$ if $T_0$ is high (orange trajectory) and to increase $T$ if $T_0$ is low (blue trajectory). 
    Then, when $\sigma \gtrsim 0.6 $, we decrease the temperature \UL{ and keep it around $T\in[0.4,0.5]$ to trap the particles kinetically.} 
    \UL{As a result, the structures remains metastable with $\sigma \approx \sigma^*=0.65$.}
    
\begin{figure}[h!]
        \centering
        \includegraphics[width=0.48\textwidth]{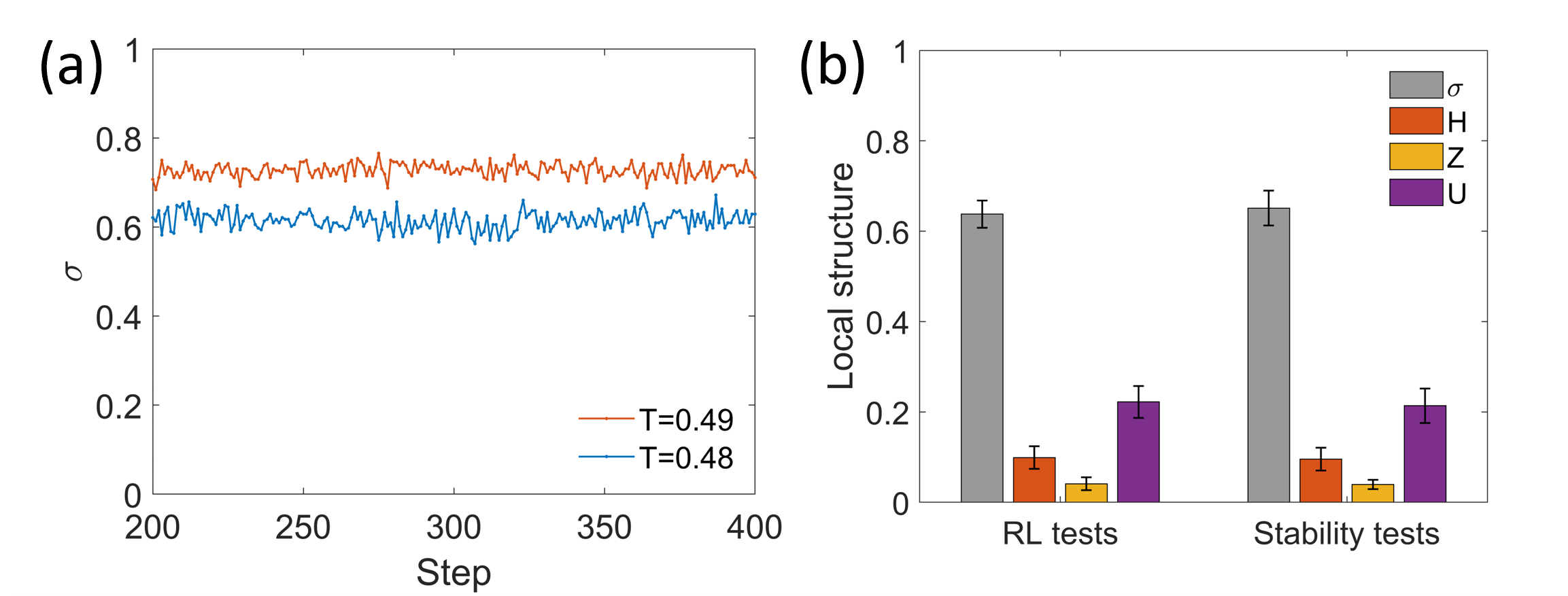}
        \caption{\UL{Stability of the RL tests for $\sigma^*=0.65$. (a) Continuation of the optimised structures from the step 200 in Fig.~\ref{fig:patchy.othertarget}(a-c) at fixed temperature. The number of BD steps is $20\times10^6$ steps, equivalent to 200 RL steps. (b) Statistics of the local structure for 20 independent samples.}}
        \label{fig:stabilise.sig065}
    \end{figure}

    \UL{Moreover, the structures obtained at the end of RL undergo stabilisation at the corresponding temperatures (Fig.~\ref{fig:stabilise.sig065}). The continuation of the two trajectories in Fig.~\ref{fig:patchy.othertarget}(c) is shown in Fig.~\ref{fig:stabilise.sig065}(a). Here the temperature is fixed at the last temperature of RL tests, which is $\approx 0.5$.
    No significant change of $\sigma$ is observed. Figure~\ref{fig:stabilise.sig065}(b) indicates that $\sigma$ is statistically maintained around $\sigma^*=0.65$ before and after stabilisation. When $\sigma>0.8$,} 
    the DDQC is the undesired structure as we set $\sigma^*=0.65$. 
    Figure~\ref{fig:patchy.othertarget}(a) also shows how the policy prevents the DDQC by increasing $T$ whenever $\sigma>0.8$. 
    \UL{It can be inferred that the system will be brought to the state of high $T$ and low $\sigma$, which locates on the upper left of the policy in Fig.~\ref{fig:patchy.othertarget}(a). Then the state-action is operated somewhat similar to the trajectory in orange. This kind of behaviour when $\sigma$ deviates from the target is observed more clearly when the target structure is $\sigma^*=0.35$.}

    \begin{figure}[!]
        \centering
        \includegraphics[width=0.40\textwidth]{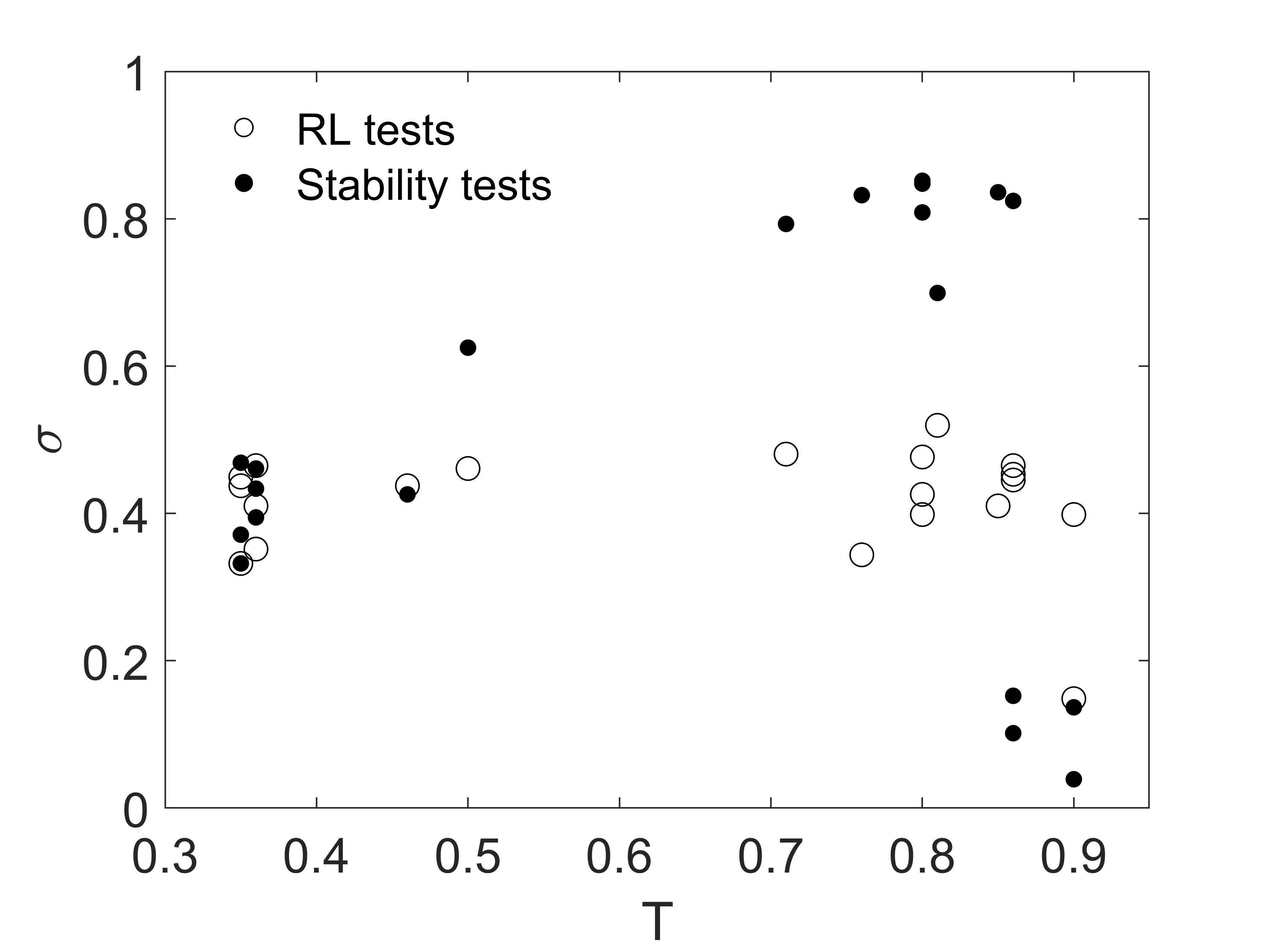}
        \caption{\UL{Stability of the RL tests for $\sigma^*=0.35$. Structures from the step 1000 in Fig.~\ref{fig:patchy.othertarget}(d-f) continue to be simulated at the corresponding temperature. The number of BD steps is $20\times10^6$ steps, equivalent to 200 RL steps.}}
        \label{fig:stabilise.sig035}
    \end{figure}   
    In Fig.~\ref{fig:patchy.othertarget}(d-f), the policy and tests for the target $\sigma^*=0.35$ are demonstrated. 
    The policy can be divided into three regimes, represented by the snapshots 1, 2, and 3. At first, large $\sigma$ structure ($\sigma>0.5$) is avoided by increasing temperature (see snapshot 1 of Fig.~\ref{fig:patchy.othertarget}(d-f)). 
    When $T\approx 0.85$, the structure strongly fluctuates with $0<\sigma< 0.65$, for example, between snapshots 1 and 2. 
    When $\sigma$ becomes small, the blue region in the policy around snapshot 2 suggests decreasing temperature, and the system attempts to reach a state such as snapshot 3. 
    The structure near snapshot 3 is not stable, and after a long time, the structure \UL{deviates from the target, i.e. $\sigma >0.5$}. 
    Then, a new cycle of snapshots $3 \rightarrow 1 \rightarrow 2$ occurs.     
    \UL{
    We check the stability of the obtained structure from the RL test, similar to the case of $\sigma^*=0.91$ and $\sigma^*=0.65$.
    We fix the temperature after the RL tests at the corresponding temperature after 1000 RL steps (Fig.~\ref{fig:stabilise.sig035}). The samples before fixing the temperature have $\sigma$ close to $\sigma^*$. 
    Figure \ref{fig:stabilise.sig035} shows the results of the stability test in comparison with the last structure of the RL test. 
    The results show that the obtained structure is not stable for the target $\sigma^*=0.35$.
    After fixing the temperature, some tests at $T<0.5$ still have their $\sigma$ fluctuate around $\sigma^*$, while some deviate from the target. In order to drive those tests to the target, the temperature should follow the policy.}
    The result reveals that RL can learn even when the target is unstable. The policy shows how we can obtain the target structure dynamically by changing the temperature.

    \subsection{\NY{RL, equilibrium phases, and metastability}}
    \label{sec.REMC}    
    
     We have investigated how RL agent learns and proposes policies for temperature control of patchy particles to form a DDQC.
    Our results suggest that the best policy for making DDQC is to change the temperature quickly to \UL{a characteristic} temperature $T^*=0.7$, keep the temperature until the system is dominated by the dodecagonal structures, and then decrease the temperature further to get the DDQC \UL{stabilised}. It is noted that the \UL{characteristic temperature} $T^*$ is autonomously found out by RL. 
    At this temperature, the structural fluctuations are enhanced.
    As a result, there is more chance of getting the dodecagonal structure. 
    In the estimated policy by RL, if the temperature is high, the particles are too mobile to make an order structure. Hence, a decrease in temperature is suggested. 
    When we start from low $T_0$, the policy suggests increasing temperature so that the system may escape from the metastable state.

\begin{figure}[!]
        \centering
        \includegraphics[width=0.4\textwidth]{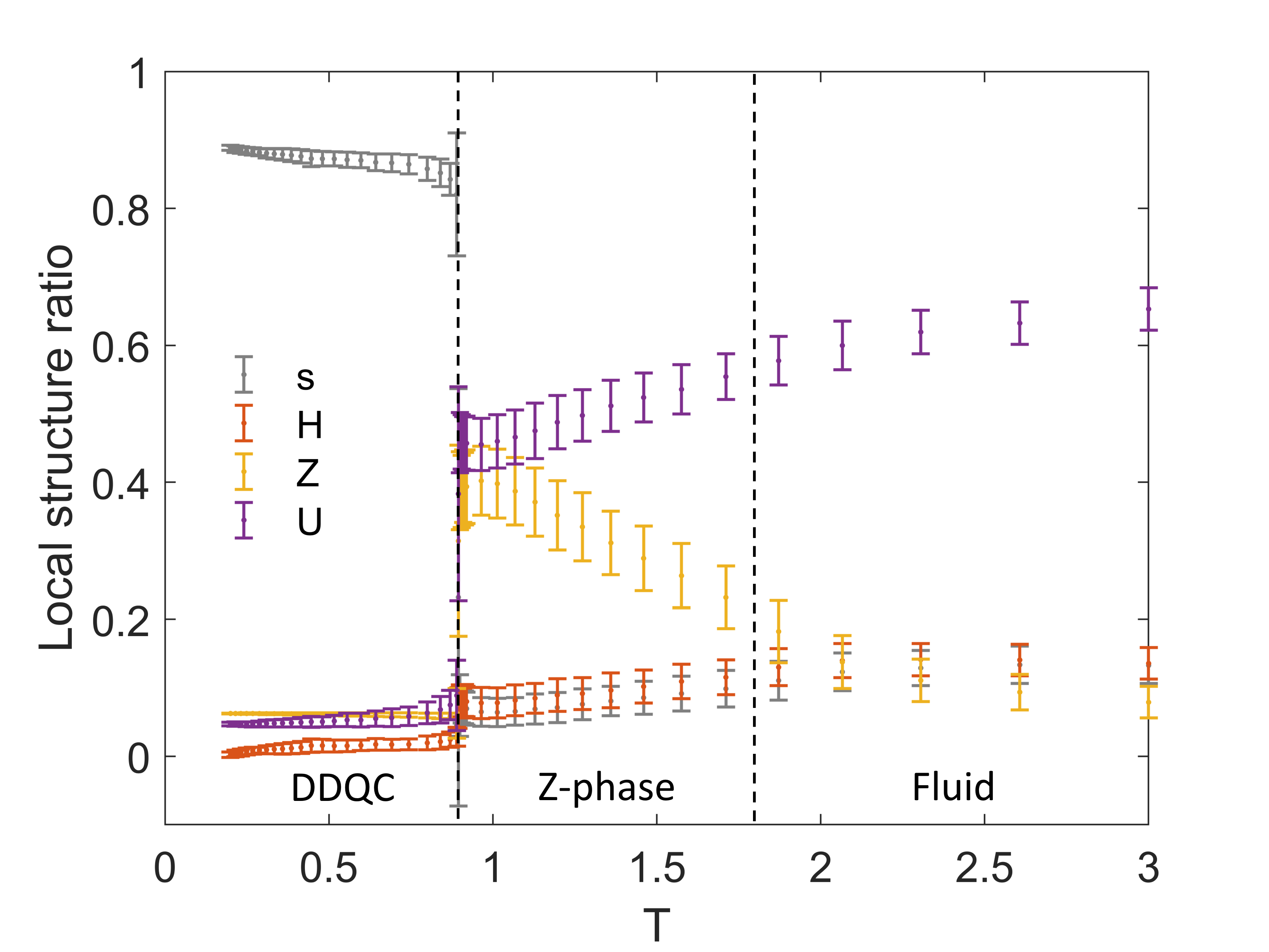}
        \caption{\UL{Dependence of the local structure on temperature, and the dominant phase by REMC. The mean (dot) and standard deviation (bar) are calculated from RE step 200 to 1000 in Fig.~\ref{fig:S.REMC.temperature}.}}
        \label{fig:REMC}
    \end{figure}

    \UL{Our RL suggests that the policy changes at the characteristic temperature $T^*=0.7$.
    \NY{Figure \ref{fig:REMC} shows an equilibrium local structure at each temperature computed by REMC.
    The phase transition between $Z$-phase and DDQC occurs at $T\approx 0.89$ (see Sec.~\ref{secS:REMC}). 
    }
    Simulations of random initial configurations with finite cooling rate methods such as quenching (rapid temperature change) and annealing (slow temperature change) show the transition at lower temperatures. 
    In quenching (Fig.~\ref{fig:S.quenching}), the DDQC is formed when $T\approx[0.7,0.85]$. 
    If $T<0.7$, the patchy particles cannot form DDQCs because the system gets trapped in the metastable states, and it is unlikely to remove the defects \NY{at the fixed temperature}.
    In this sense, the characteristic temperature $T^*=0.7$ coincides with the lower limit of fixed $T$ setting for DDQC. 
    On the other hand, during the annealing (Fig.~\ref{fig:RL.vs.annealing}), the system has more chance to escape from the metastable states. The change of phase is observed at $T\approx 0.8$. 
    Because of the nature of the temperature change at finite speed in RL, the transition temperature becomes effectively lower in RL.
    We should note that in the policy of RL, the temperature is discretised in the mesh size of 0.1.
     $T^*$ may exhibit deviations dependent on the mesh size. Another point to be noted is that $T^*$ in the policy exists when the current structure is not a DDQC, in particular $\sigma < 0.7$. When a DDQC structure is obtained, the temperature should be kept around 0.5.
}

 \NY{We should emphasise that our RL does not optimise the distribution of $\sigma$ as a function of temperature, nor the phase diagram. The RL policy suggests the most rewarding pathway to reach the target.
    }
    RL can learn that the \UL{characteristic} temperature $T^*$ plays an important role in enhancing the probability of QC structural formation. 
    RL method automatically finds them during the training steps. The method feeds neither existence of the this temperature nor its value. 

\section{Discussion and conclusion} \label{sec.conclusions}

   \NY{Before summarising our study, we discuss several issues to clarify the mechanism and generalisation of RL and to compare our RL with other studies.
   } 
    In order to show how RL works to overcome the energy barriers between the disordered, metastable and DDQC structures, we consider a point particle in a temperature-dependent triple-well potential.
    \UL{The potential is designed to mimic the disorder, metastable, and global minimum state.} Analogically, a characteristic temperature is also found to avoid the metastable state while increase the chance toward global minimum (see Sec.~\ref{secS:triple.well}).

    

    Q-learning RL in this study can be applied to various self-assembly systems.
    We demonstrate it for the system of patchy particles and isotropically interacting particles (see Sec.~\ref{secS:iso}).
    Both systems show DDQC; nevertheless, the estimated policy of temperature control is qualitatively different.
    \UL{The policy in an isotropic system is much simpler, as no characteristic temperature is found.}
    The results give us physical insights on the two systems.
    The system of patchy particles has metastable states, which have to be overcome to form DDQC, whereas the system of isotropically interacting particles is monostable.

    We focus on the estimation of the policy for DDQC, which is stable at a certain range of temperature. However, our RL method is not limited to such a stable target structure. In fact, we demonstrate that RL can estimate temperature protocol for metastable and even transiently stable structures. 
    To obtain the metastable structure as a target ($\sigma^*=0.65$), we may leave the system near the \UL{characteristic} temperature \UL{$T^*=0.7$} at which structural fluctuation is large. Then rapidly decrease the temperature so that the structure is frozen at the desired metastable state. 
    When the target is not even at the metastable state ($\sigma^*=0.35$), the policy suggests to wait at a certain low temperature to obtain the target structure. In this case, the structure is transient, and after some time, it escapes from the target. Then the temperature changes so that the system returns to the target structure.

    Those results, including the temperature protocol for DDQC, may be reached from sophisticated guess, but we think this is not the case for many people.
    We believe that RL, like any machine learning method, can assist our finding mechanisms of unknown phenomena and making decisions more efficiently. To tackle more complex, highly non-linear, and high dimensional problems, the combination of machine learning with expertise in decision making may help to understand the problem better.

    The choice of statistical quantities that characterise the structures is crucial for designing a successful RL system. This includes the choice of the relevant states and how finely to discretise the states (for Q-table). In the case of DDQC, the continuous state we choose is the ratio of the $\sigma$ particles because $\sigma$ can span over a wide range in $[0,1]$ under the investigated temperature. Therefore, the states can distinguish the DDQC from metastable and disordered structures.
    One can consider the $Z$ particles to evaluate the DDQC structure. However, under the same condition, the performance of Q-learning with $Z$ is not as good as Q-learning with $\sigma$ because the ratio of $Z$ spans over a much narrower range. 
    Methodologically, there is no limit of number of states in RL. For example, one may include two microscopic states, e.g. $\sigma$ and $Z$. When the dimension of states is much higher, the computational cost using Q-table is too high. Approximation of the Q-function by the small number of continuous basis functions is promising in this direction.

    In this study, we use the states of $T$ and $\sigma$, the action space of change in temperature $\Delta T$, and the reward function of $(\sigma -\sigma^*)^2$. However, we still need to consider many hyperparameters, such as the number of epochs $N_\text{e}$ during training and the effect of discretisation. We discuss some general issues: how prior knowledge can help reduce the calculation cost, the effect of discretisation of Q-table, and the effect of $\epsilon$-greedy, in Supplementary Information.

    There are many ways of doing reinforcement learning.\cite{brunton_2022,ravichandiran_2020} In their study on RL for self-assembly,\cite{Whitelam:2020} Whitelam and Tamblyn have shown that the evolutionary optimisation to train the neural network can learn actions on the control parameters, such as temperature and chemical potential, for the self-assembly of a target structure. 
    Evolutionary optimisation takes a black-box approach to learn the action as a function of the state (or time), which is expressed by the weights in the neural network.\cite{Salimans:2017}
    On the other hand, Q-learning relies on the maximisation of future reward, which is expressed by the Bellman's equation. 
    The sampling during training is also different in the two methods. 
    The evolutionary optimisation requires the final outcome of the trajectory of the self-assembly process, while the Q-learning updates the policy iteratively by observing the state-action pair during the dynamical process. 
    As a result, Q-learning works on-the-fly and requires less computational cost compared to evolutionary optimisation. 
    We should stress that regardless of the differences, both evolution-type optimisation and Q-learning based on the Markov decision process estimate the policy that can produce the target faster than a conventional cooling scheme. 
    More studies are necessary to clarify generic guidelines on how to choose a suitable RL model.

    Although RL can estimate the best temperature protocol, it has to be related with the physical properties of the system. 
    The work in Ref.\cite{bupathy_2022} proposed a temperature protocol based on free energy calculation of nucleation barrier and metastability of the free energy minima. Although it treated a toy model, relating the physical properties of QC formation and performance of RL would be an interesting future direction.

    To summarise, we propose the method based on RL to estimate the best policy of temperature control for the self-assemblies of patchy particles to obtain the DDQC structures.
    From the estimated policy, we successfully obtain the DDQCs even for the system size larger than the size we use for training.
    The key to the success is that RL finds the \UL{characteristic} temperature of the DDQC self-assembly during training.
    The estimated policy suggests that first, we change the temperature to the \UL{characteristic} temperature so that the larger fluctuations enhance the probability of forming DDQC, and then decrease the temperature slightly to remove defects.
    The mechanism of learning optimal policy is demonstrated in the simple triple-well model.
    In order to avoid metastable states, the optimal policy suggests increasing the temperature if we start from a low temperature. 
    The RL is capable of giving insights to different self-assembled systems, and dynamically adapting the policy in response to unstable target. 
    We should stress that our method can be applied to other parameters that we may control.
    Therefore, we believe that the method presented in this work can be applied to other self-assembly problems.

\section*{Author Contributions}
U.L. performed the simulations and analysed the data. N.Y. designed the research. All the authors developed the method and were involved in the evaluation of the data and the preparation of the manuscript.

\section*{Conflicts of interest}
There are no conflicts to declare.

\section*{Data Availability}
The data supporting this article have been included as part of the Supplementary Information. The codes of RL for self-assembly of patchy particles and RL for triple-well model can be found at \url{https://github.com/ULieu/RL_patchy} and \url{https://github.com/ULieu/RL_3well}.

\section*{Acknowledgements}
The authors acknowledge the support from JSPS KAKENHI Grant number JP20K14437, JP23K13078 to U.L., and JP20K03874 to N.Y. This work is support also by JST FOREST Program Grant Number
JPMJFR2140 to N.Y.
The authors would like to thank Rafael A. Monteiro for bringing the idea of reinforcement learning to our attention. 






\bibliography{RLpaper} 
\bibliographystyle{rsc} 
\end{document}


\title{Dynamic Control of Self-assembly of Quasicrystalline Structures through Reinforcement Learning\\
SUPPLEMENTARY INFORMATION}

\author{Uyen Tu Lieu$^{\ast}$\textit{$^{a,b}$} and Natsuhiko Yoshinaga\textit{$^{a,b}$}}
\date{}

\maketitle

\textit{$^{a}$Future University Hakodate, Kamedanakano-cho 116-2, Hokkaido 041-8655, Japan}

\textit{$^{b}$Mathematics for Advanced Materials-OIL, AIST, 2-1-1 Katahira, Aoba, 980-8577 Sendai, Japan}

\textit{E-mail: uyenlieu@fun.ac.jp; yoshinaga@fun.ac.jp}

\renewcommand{\thefigure}{S\arabic{figure}}
\renewcommand{\thetable}{S\arabic{table}}
\renewcommand{\theequation}{S\arabic{equation}}
\renewcommand{\thesection}{S\arabic{section}}

\section{Reinforcement learning and Q-learning} \label{secS:RL.and.Qlearning}

    The basic ingredients of reinforcement learning (RL) include an agent, an environment, and reward signals. 
    The agent observes the states $s$ of the environment and learns optimal actions $a$ through a policy $\pi$ that maximises the cumulative future rewards $R$ \cite{brunton_2022,sutton_1998,ravichandiran_2020}.
    The future reward is the sum of the instantaneous reward $r_i$ at each step $i$ 
        \begin{align}
            R
            &=
            \sum_i \gamma r_i
        ,
        \end{align}
    with the discount factor $\gamma$.
    Formally, RL is expressed by a tuple of $\{ \mathcal{S}, \mathcal{A}, \mathcal{P}, g, \pi \}$ where $\mathcal{S}$ is a state space, $\mathcal{A}$ is an action space, $\mathcal{P}$ is a Markov transition process of the environment describing its time evolution, $g$ is the (instantaneous) reward function, and $\pi$ is a policy.
    The transition process $P (s_{t+1} | s_t, a_t) \in \mathcal{P}$ maps the current state $s_t \in \mathcal{S}$ to the next state $s_{t+1}$ under the action $a_t \in \mathcal{A}$.
    The process is supplemented by the initial probability of the states $P (s_0)$.
    The reward measures whether the current state is good or bad. 
    The reward function gives some numbers from the current state and action as $r_t = g(s_t, a_t)$.
    In this work, we assume the reward function is dependent only on the state, that is, $r_t = g(s_t)$.
    In general, the policy is a conditional probability $\pi (a|s)$ of taking the action under a given state.
    We assume the deterministic policy $a(s)$, namely, the action is the function of the state.

    A physical interpretation of RL is to estimate the best dynamic control strategy to get a desired structure or physical property. 
    The physical system of variables $s_t$ yields the dynamics expressed by the Markov process $\mathcal{P} (s_{t+1} | s_t, a_t)$ under an external force and/or parameter change in the model expressed by $a_t$.
    At each time, we can compare the current state $s_t$ and the target state $s^*$.
    The distance between them is an instantaneous reward.
    The goal of RL is to estimate the best policy from which we choose the action $a$ as a function of the current state $s$.

    There are many RL algorithms to train the agent. Q-learning is a popular algorithm for learning optimal policies in Markov decision processes \cite{sutton_1998}. It is a model-free, value-based algorithm that uses the concept of Q-values (Quality value) to guide the agent's decision-making process. Q-value, denoted as $Q(s,a)$, is the cumulative reward obtained by taking action $a$ on the current state $s$ and then following the optimal policy.  
    The simplest Q-learning uses a Q-table in which Q-values are updated at each point in discretised action and state spaces. The size of Q-table depends on the number of elements of the state spaces and action spaces. For example, consider a system with two state spaces discretised into $m$ and $n$ elements, and an action space with $l$ elements. In this case, the corresponding Q-table is a three-dimensional array with dimensions $m\times n \times l$. This array represents the whole state-action space, in which the agent (we) can store and update Q-values for all possible combinations of states and actions. The policy is then extracted from the Q-value of each state-action pair $Q(s,a)$. In general, the algorithm involves many epochs (or episodes). The Q-table is initialised at first. 
    For each epoch, the states are also initialised, then for each step in the epoch, we perform the following algorithms:
    \begin{itemize}
        \item Observe a current state $s_t  \in \mathcal{S}$.   
        \item Select and perform an action $a_t \in \mathcal{A}$ based on the policy from $Q(s,a)$.   
        \item Observe the subsequent state $s_{t+1}$.    
        \item Receive an immediate reward $r_{t+1}$. 
        \item Update iteratively the Q-function by
    \end{itemize}
    
    \begin{equation}
        \label{eq:Qupdate}
        Q(s_t,a_t)=Q(s_t,a_t)+\alpha[r_{t+1} + \gamma \max_{a} Q(s_{t+1},a)- Q(s_t,a_t)]
    ,
    \end{equation}
    where the learning rate $\alpha$ is a hyperparameter $0\leq \alpha \leq 1$ that reflects the magnitude of the change to $Q(s_t,a_t)$ and the extent that the new information overrides the old information. If $\alpha=0$, no update at all; if $\alpha=1$, then completely new information is updated in $Q$. The discount factor $\gamma$ is associated with future uncertainty or the importance of the future rewards $(0\leq \gamma \leq 1)$.

    In RL, it is important to consider the balance between exploitation and exploration. 
    If we just follow the current (non-optimal) policy, it is unlikely to find potentially more desired states.
    On the other hand, if our search is merely random, it takes a significant amount of time to find them. In Q-learning, exploitation involves selecting the action that is believed to be optimal, i.e. maximum Q-value, while exploration involves selecting the action that does not need to be optimal within the current knowledge. To balance these strategies, the $\epsilon$-greedy method is used. In the $\epsilon$-greedy method, a random action at each time step is selected with a fixed probability $0\leq \epsilon \leq 1$ instead of the optimal action with respect to the Q-table.
    \begin{equation}
        \pi(s)=
        \begin{cases}
            \text{random action } a\in \mathcal{A}, 
                & \text{if } \xi < \epsilon    \\
            \underset{a\in \mathcal{A}}{\arg\max} Q(s,a),& \text{otherwise,}
        \end{cases}
    \end{equation} 
    where $0\leq \xi \leq 1$ is a uniform random number at each step.

\section{Reinforcement learning with value iteration method} \label{secS:value.iteration}

    Q-learning is a model-free method in RL. As shown in the main article, the Q-table is updated during training (BD simulations at given $T$) and eventually the policy is determined from the Q-table. Here we propose to use value iteration to utilise the data from training Q-table. Value iteration is a model-based method, i.e. we need to know the model dynamics (transition probability of the next states given current states and actions) \cite{brunton_2022,ravichandiran_2020}. 
    Therefore, we first estimate the transition probability from the current state $s=(\sigma, T)$ to the next state $s'$ under the action $a$, $P(\sigma', T'| \sigma, T, a)$, from empirical sampling. 
    Then, we use Bellman's equation to estimate the value function, from which we can estimate the policy of the temperature change.
    Here, we show how to calculate the value function
    
    \begin{enumerate}
        \item Sampling data $S(s,a,s')$ 
        \item Discretising the state spaces and calculating the transition probability $P(s'|s,a)$
        \item Performing value iteration on the discretised state space: 
        \begin{itemize}
            \item initialise the value function $V(s)=0$
            \item in each iteration, calculate 
                \begin{equation}
                Q(s,a)=\sum_{s'} P(s'|s,a) [R(s,a,s')+\gamma V(s')];
                \end{equation}
        \end{itemize}
    \end{enumerate}   
    and the value function in this iteration is $V(s)=\max_a Q(s,a)$. 
    
    For the self-assembly of patchy particles in our study, there are two states $s=(\sigma,T)$. We report the result after 100 iterations when the value function converges. Note that the calculation of $V(\sigma,T)$ and policy uses the information of $P(\sigma', T'|\sigma,T, a)$. 
    The hyperparameters such as the target $\sigma^*$, reward function, can also be varied. Once the value function converges, one can determine the corresponding Q-value $Q(\sigma,T,a)$ and the policy 
    $\pi(a|\sigma,T)=\arg \max_a Q(\sigma,T,a)$. 
    The reward function and discount factor 
    $\gamma$ in value iteration are chosen identical to that in Q-learning.

\section{\UL{Replica Exchange Monte Carlo simulations}}\label{secS:REMC}

    Replica Exchange Monte Carlo method (REMC, also called parallel tempering) \cite{sugita_2000,iba_2001} is employed to investigate the equilibrium phase diagram of the five-fold patchy particles used in the study. In REMC, $N_\text{rep}$ noninteracting replicas are simulated simultaneously at a range of inverse temperature $\beta_1 \leq \beta_2 \leq ...\leq \beta_{N_\text{rep}}$ by standard Monte Carlo (MC) simulations \cite{allen_2017}. After a fixed number of Monte Carlo sweeps, extra replica exchanges (swapping) between replicas of neighbouring parameter $\beta_i$ and $\beta_{i+1}$ is suggested and accepted with a probability
    \begin{equation}
        p(E_i,\beta_i \rightarrow E_{i+1}, \beta_{i+1})=\min(1,\exp( (\beta_i-\beta_{i+1}) (E(\mathbf{x}_i)-E(\mathbf{x}_{i+1})),
    \end{equation}
    where $\mathbf{x}$ is the particle configuration. 
      
    In the study, we set the distribution of inverse temperature $\{ \beta_i \}$ on replicas based on the distribution of energy $E(\mathbf{x},\beta)$. In order to facilitate the exchange, we chose $\beta$ so that $(\beta_i-\beta_{i+1})(E(\mathbf{x}_i)-E(\mathbf{x}_{i+1}) \sim \mathcal{O}(1)$. 
    We use the same initial condition for all replicas.
    Initially, the configuration of each replica evolves under its temperature.
    When exchange occurs at each REMC step, the temperatures between the neighbouring temperatures are swapped, i.e. $\left( \beta_i, \beta_{i+1} \right) \rightarrow \left( \beta_{i+1}, \beta_{i} \right)$. The simulation condition for REMC simulation is given in Table~\ref{table:REMC}.
    
    \begin{table}[h]
        \centering
        \small
        \setlength{\tabcolsep}{3pt}
        \begin{tabular}{|l|c|}
            \hline
        Parameter & Value          \\      
            \hline
        Potential & Patchy particle\\
        Number of particles, $N$  & 256\\
        Area fraction & 0.75\\
        Initial configuration & DDQC structure\\   Number of replicas, $N_\text{rep}$ & 44\\
        Temperature & $[0.2,3.0]$\\   
        Number of MC sweeps per replica exchange (RE) step& 100,000\\
        Number of replica exchange steps & 1,000\\
            \hline   
        \end{tabular}
        \caption{    
        Setting of REMC simulations. Details of temperatures can be seen in Fig.~\ref{fig:S.REMC.temperature}-\ref{fig:S.REMC.energy}. 
        }
        \label{table:REMC}    
    \end{table}   

\begin{figure}[!]
        \centering
        \includegraphics[width=0.7\textwidth]{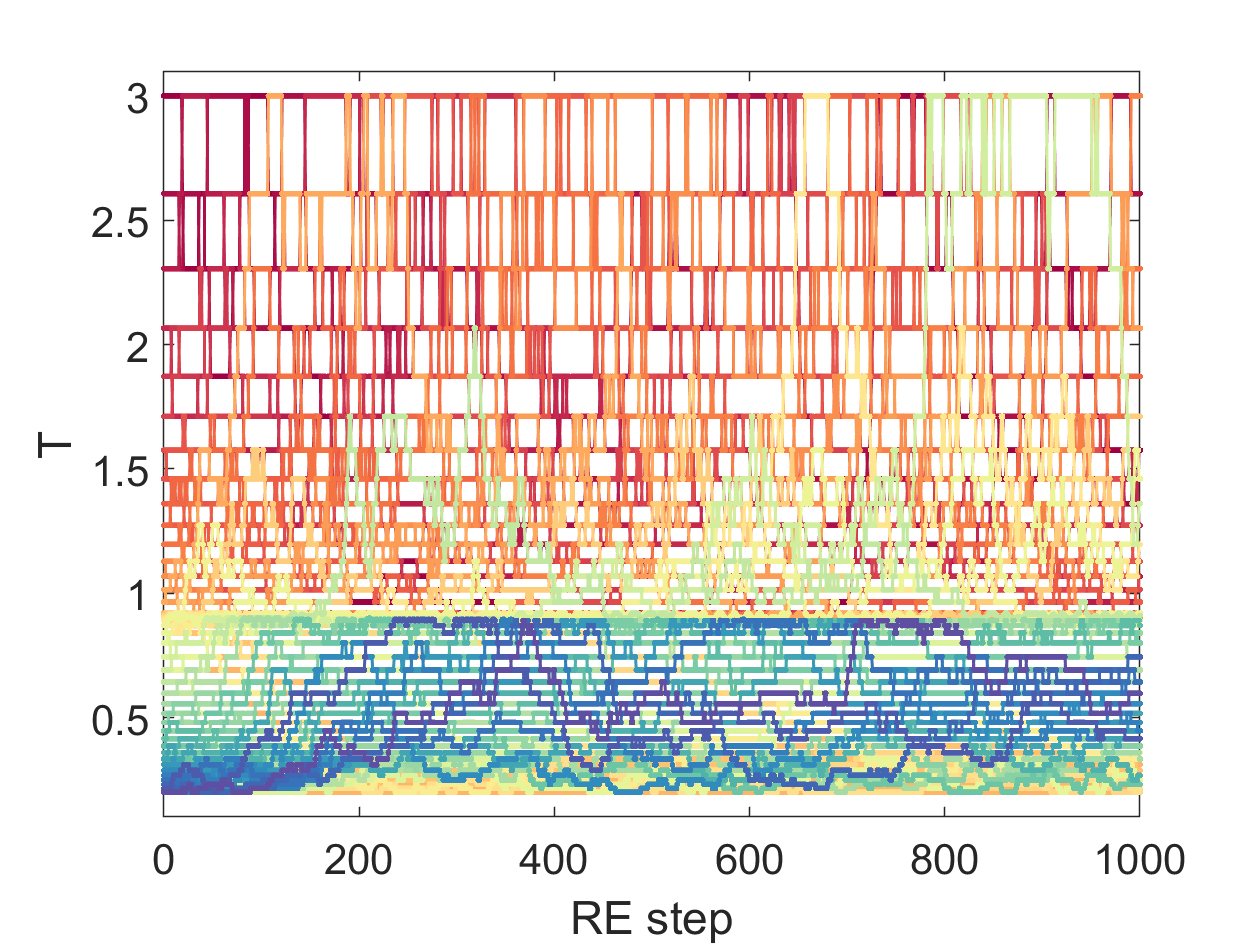}
        \caption{Temperature profiles of replica exchange Monte Carlo simulations for the system of $N=256$, $A=0.75$ (Table~\ref{table:REMC}). Colour corresponds to the replica index.}
        \label{fig:S.REMC.temperature}
    \end{figure}

\begin{figure}[!]
        \centering
        \includegraphics[width=0.8\textwidth]{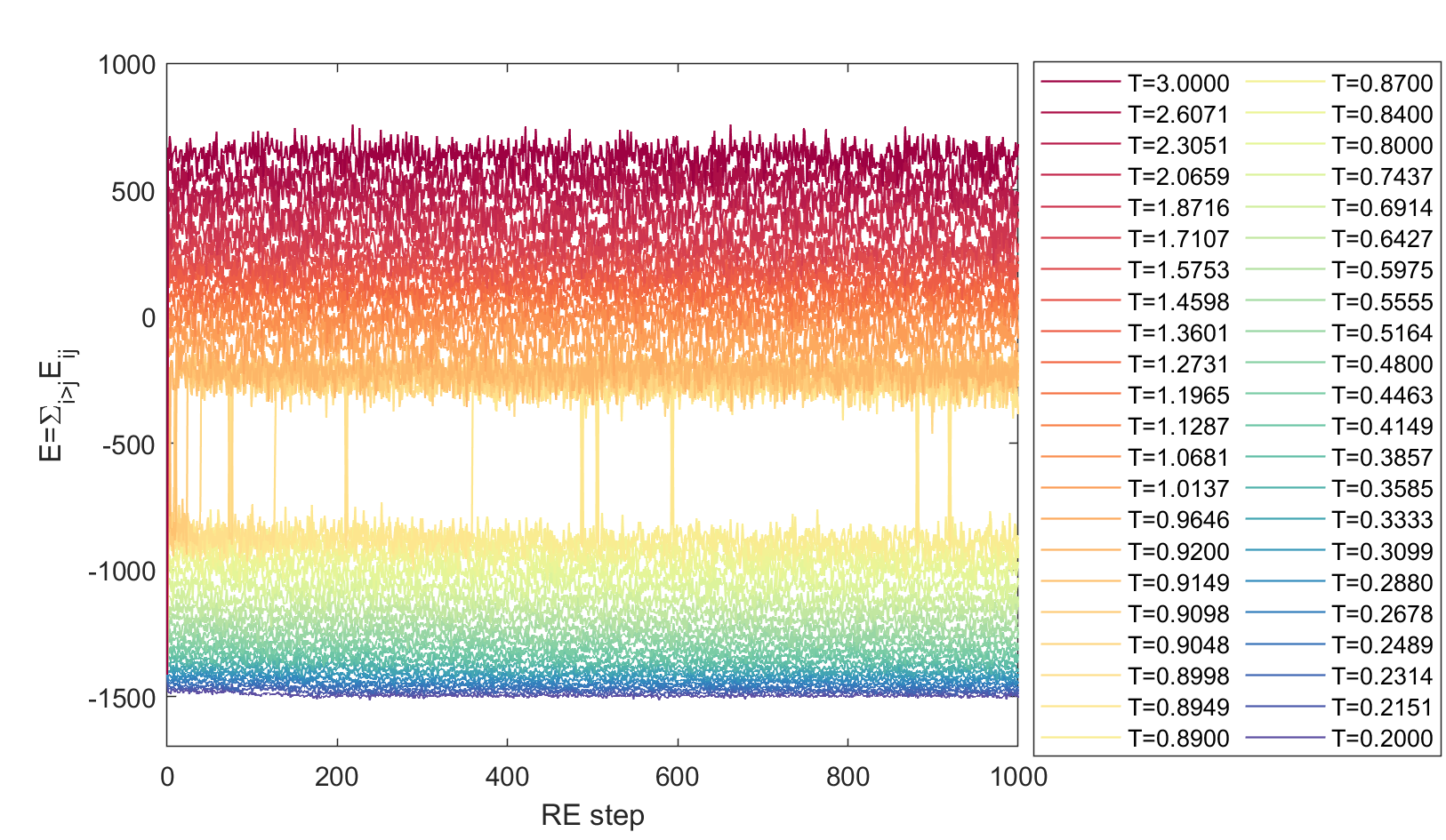}
        \caption{Energy profiles for the replica exchange system in Table~\ref{table:REMC}. Colour corresponds to temperature.}
        \label{fig:S.REMC.energy}
    \end{figure}

\begin{figure}[!]
        \centering
        \includegraphics[width=0.9\textwidth]{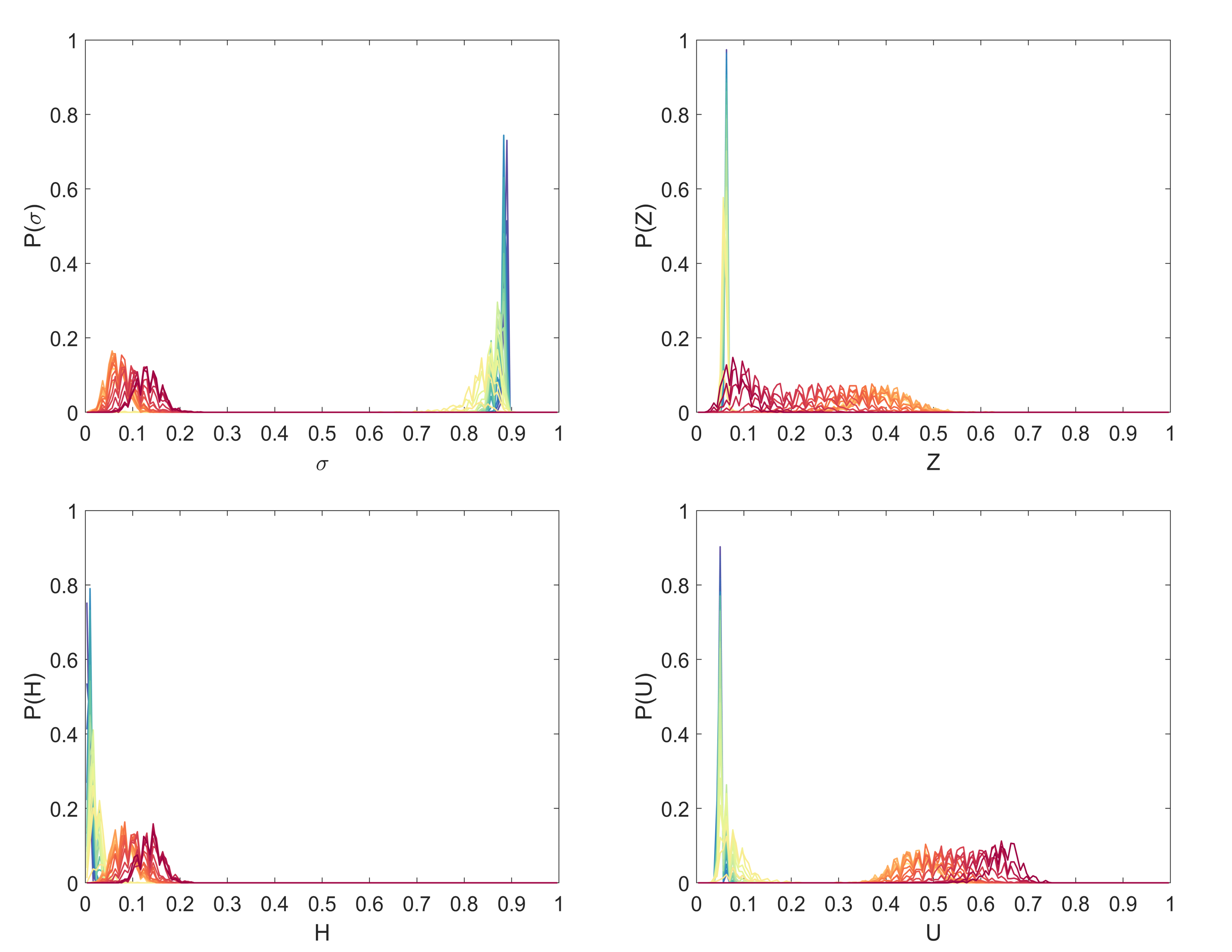}
        \caption{Probability distributions of the ratios of local structure for the replica exchange system in Table~\ref{table:REMC}. The data is taken from RL step 200 to 1000. Colour corresponds to temperature like Fig.~\ref{fig:S.REMC.energy}.}
        \label{fig:S.REMC.dist.local}
    \end{figure}

\begin{figure}[!]
        \centering
        \includegraphics[width=0.7\textwidth]{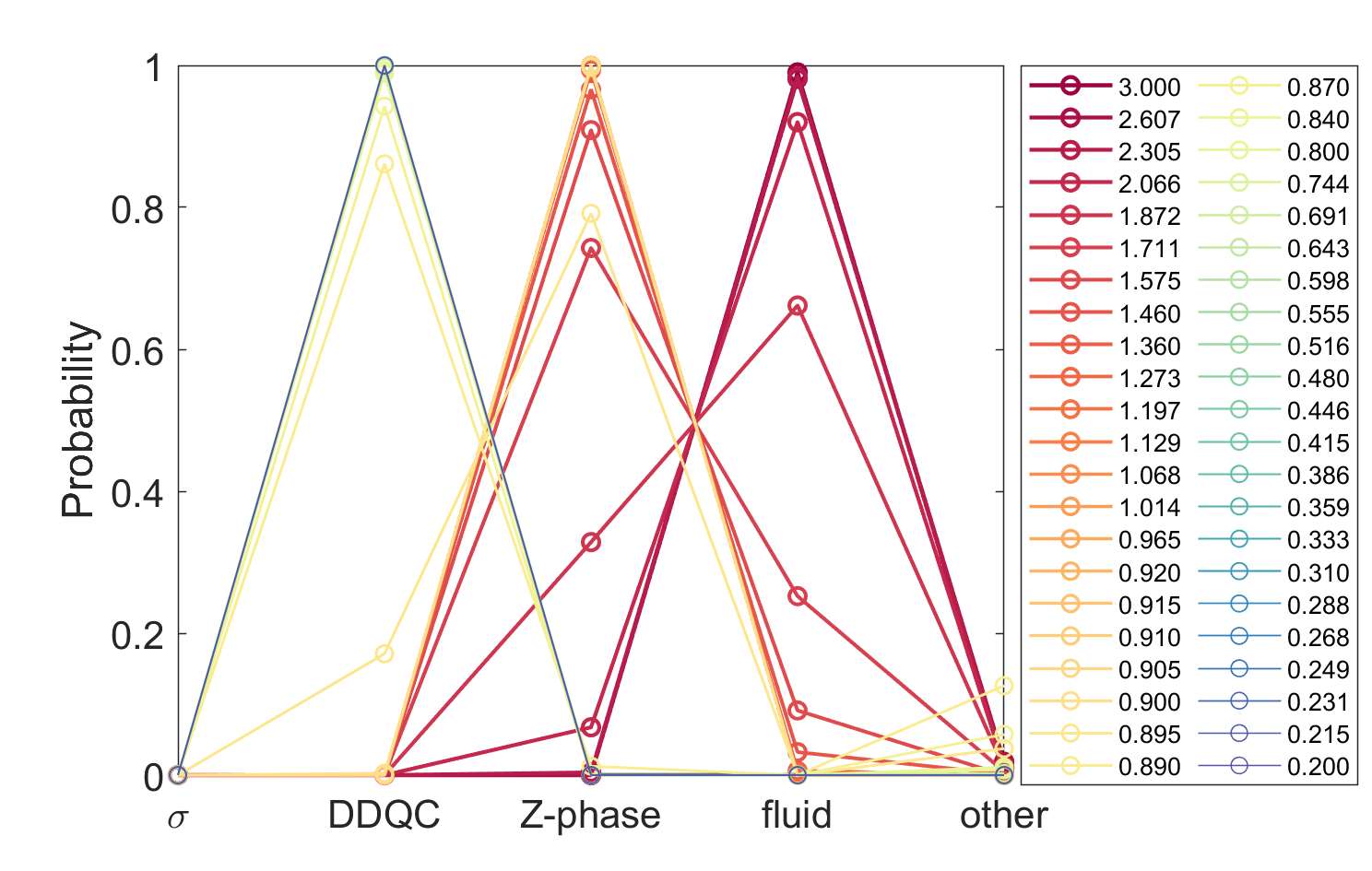}
        \caption{Probability distributions of the local structure ratios for the replica exchange system in Table~\ref{table:REMC}. Colour corresponds to temperature like Fig.~\ref{fig:S.REMC.energy}. The unknown state `other' occupies only a small portion, thus brings no major effect to the phase diagram.}
        \label{fig:S.REMC.dist.state}
    \end{figure}

    The profile of the temperature of each replica index during REMC simulations is given in Fig.~\ref{fig:S.REMC.temperature}. The energy corresponding to the temperature in Fig.~\ref{fig:S.REMC.energy} shows a huge gap at $T>0.9$ and $T<0.85$, indicating a phase transition. We determine the ratio of the local structure $\sigma$, $Z$, $H$, $U$ (undefined) at each REMC step. The probability distributions of those ratios are depicted in Fig.~\ref{fig:S.REMC.dist.local}. In order to define the state of the self-assembly, e.g. dodecagonal quasicrystal (DDQC), $\sigma$ phase, etc., we use the criteria based on the local structures (Table~\ref{table:S.criteria.phase}). Figure~\ref{fig:S.REMC.dist.state} illustrates the probability distribution of these states as a function of temperature. We assume that the state with the highest probability is the dominant state. 
    There are three main phases: DDQC, Z-phase and fluid. The phase transition between the Z-phase and DDQC corresponds to the huge energy gap in Fig.~\ref{fig:S.REMC.energy}. For the temperature  $ T\in [0.2,1.3]$ used in the reinforcement learning, the phases are DDQC and Z-rich.
    The dependence of the local structure on temperature with the phase is presented in the main text. 
   
    \begin{table}[h]
        \centering
        \small
        \setlength{\tabcolsep}{3pt}
        \setlength\extrarowheight{0pt} 
        \begin{tabular}{|c|c|c|c|}
            \hline
                State   
                & $\sigma$                  
                & Z                
                & H     \\      
            \hline
                $\sigma$ phase   
                & $\sigma \geq 0.93$  
                & $Z<0.02$     
                &   \\
            DDQC        
            & $ 0.8 \leq \sigma < 0.93$  
            & $0.04\leq Z \leq0.12$   
            &   \\
                Z-phase     
                & $\sigma < 0.2$  
                & $Z \geq 0.2$    
                & \\
            Fluid      
            & $\sigma < 0.2$ 
            &    $Z < 0.2$    
            & $H<0.2$   \\            
            \hline   
        \end{tabular}
        \caption{Criteria for the structures in REMC. The Fourier transformations of the DDQC, Z-phase snapshots have 12-fold symmetric spots, triangulated spots, respectively. When the assembly does not belong to any state, we name it as `other'.}
        \label{table:S.criteria.phase}    
    \end{table}

\section{\UL{Quenching simulations at fixed temperatures}}\label{secS:quenching}
\begin{figure}[!]
        \centering
        \includegraphics[width=0.6\textwidth]{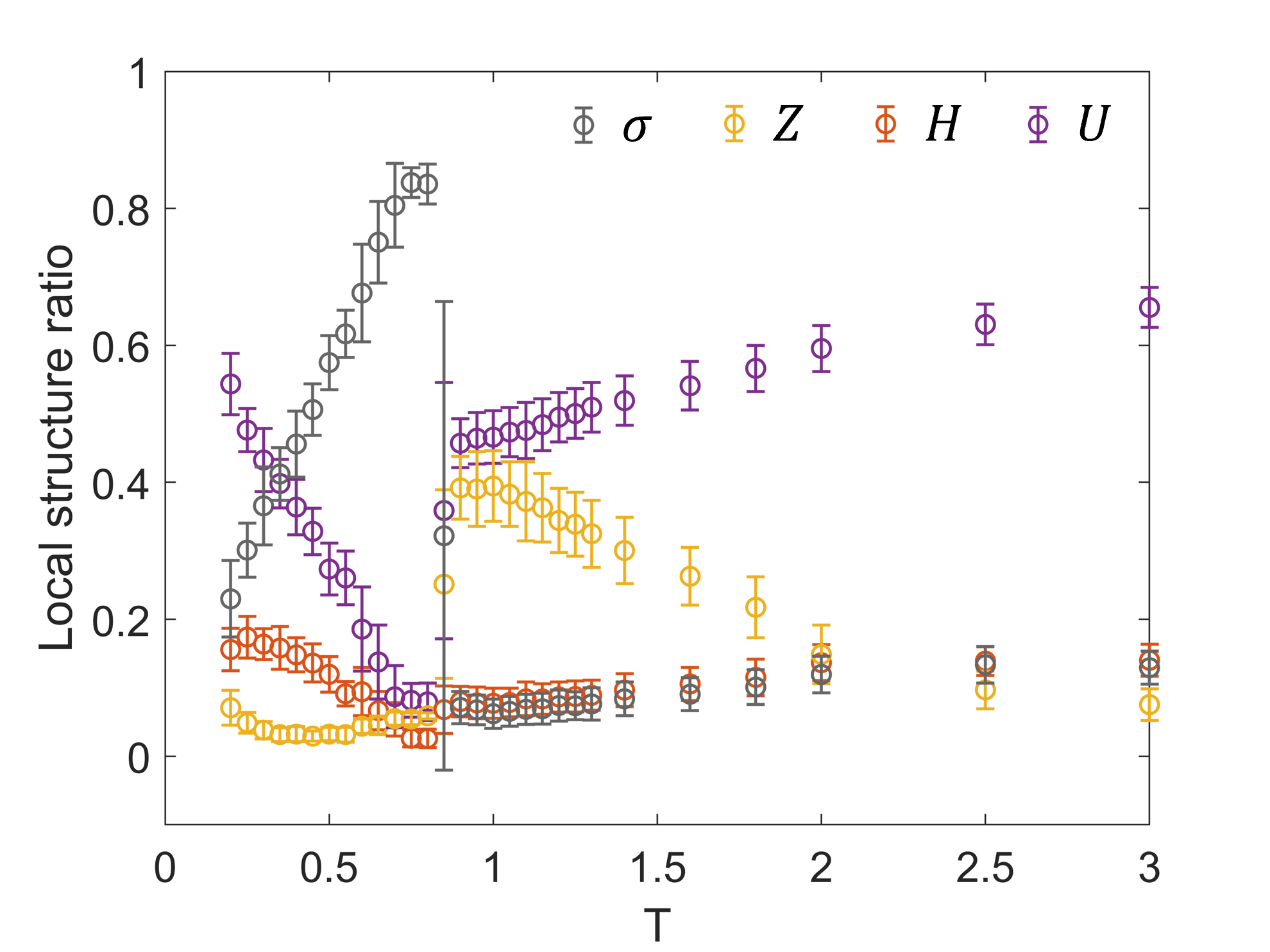}
        \caption{Assemblies by Brownian dynamics simulation at fixed temperature. The mean (circle) and standard deviation (bar) of the local structures are calculated over 10 independent samples.}
        \label{fig:S.quenching}
    \end{figure}

    Brownian dynamics simulations of patchy particles at fixed temperature setting are performed at the conditions similar to RL test. The initial particle configurations are randomly assigned. Temperature is fixed during simulation.  The number of particles is $N=256$. The spherical particles are confined in a flat plane of the size $L_x \times L_y \times 2a$ where $a$ is the particle radius and $L_x=L_y$. The periodic boundary condition is applied on $L_x$ and $L_y$. The area fraction is defined as $A=\pi a^2 N/(L_x L_y)=0.75$. The number of Brownian steps is $40\times10^6$. The dimensionless time step is set as $\Delta t=0.0005T$ when $T<2$, and $\Delta t=0.0001T$ when $T\geq 2$. The details of the local structure $\sigma$, $Z$, $H$, $U$ (undefined) at the end of simulations are given in the Fig.~\ref{fig:S.quenching}.   
    
\section{\UL{Details of local structure ratios}}    
    The details of local structure ratios for patchy particle assemblies at different temperature settings are given in Table~\ref{table:S.compare.structure}.

    \begin{table}[h]
        \centering
        \small
        \setlength{\tabcolsep}{3pt}
        \setlength\extrarowheight{0pt} 
        \begin{tabular}{|c|l|c|c|c|c|c|}
            \hline
            & Simulation conditions & $\sigma$     & H          & Z      & U      & Remarks    \\      
            \hline
        1 &$N=256$, A=0.75, RL for DDQC&$0.781 \pm 0.041$ &$0.058 \pm 0.023$ &$0.052 \pm 0.015$ &$0.109 \pm 0.035$ &DDQC\\
        2 &$N=256$, A=0.75, annealing&$0.761 \pm 0.070$ &$0.068 \pm 0.030$ &$0.055 \pm 0.009$ &$0.116 \pm 0.057$ &DDQC\\
        3 &$N=256$, A=0.75, T=0.6    &$0.637 \pm 0.066$ &$0.106 \pm 0.023$ &$0.037 \pm 0.011$ &$0.220 \pm 0.061$ &metastable\\
        4 &$N=256$, A=0.75, T=0.7    &$0.793 \pm 0.062$ &$0.049 \pm 0.027$ &$0.052 \pm 0.012$ &$0.106 \pm 0.051$ &DDQC\\    
        5 &$N=1024$, A=0.75, annealing &$0.812 \pm 0.031$ &$0.044 \pm 0.011$ &$0.061 \pm 0.003$ &$0.076 \pm 0.026$ &DDQC\\  
        6 &$N=1024$, A=0.78, annealing (Ref.\cite{lieu_2022b}) &$0.729 \pm 0.030$ &$0.076 \pm 0.011$ &$0.069 \pm 0.007$ &$0.126 \pm 0.024$ &DDQC\\
        7 &$N=256$, A=0.75, REMC at T=0.87 &$0.842 \pm 0.023$ &$0.024 \pm 0.011$ &$0.059 \pm 0.004$ &$0.075 \pm 0.021$ &DDQC\\

           \hline   
        \end{tabular}
        \caption{    
        \UL{Ratio of local structures of assemblies obtained by various temperature settings: reinforcement learning, annealing, and quenching, and REMC. The temperature in annealing is shown in the main text. There are at least 10 independent samples for the calculation of mean and standard deviation.
        }}
        \label{table:S.compare.structure}    
    \end{table}

\section{\UL{Q-learning for triple-well potential}} \label{secS:triple.well}
\subsection{Triple-well potential model}
    In RL for the DDQC self-assembly, the biggest challenge is how to avoid the metastable states and reach the global minimum state by controlling the temperature.
    In order to show how RL works to overcome the energy barriers between the disordered, metastable and DDQC structures, we consider a simple model in which a single particle at the position $x$ moves in a temperature-dependent triple-well potential.
    We design the model such that the states $x$ and $T$ correspond to $\sigma$ and $T$ of DDQC self-assembly, respectively.
 
    Similar to the DDQC, we apply Q-learning for a model consisting of two state variables $x$ and $T$.
    The states follow the dynamics described by the following Langevin equations
        \begin{align}
        x(t+dt)
        =&
        x(t)- \partial_x U(x,T) dt +\xi_x
        \label{3well.model.x}
        \\
        T(t+dt)
        =&
        T(t) + a +\xi_T
        .
        \label{3well.model.T}
        \end{align}
    The state $x$ moves in the $T$-dependent potential $U(x,T)$, whereas $T$ evolves through the action $a$ with noise.
    The functional form of the potential is shown in Fig.~\ref{fig:3well.model}.
    The fluctuation of $x$ is illustrated by the noise term $\xi_x$.
    The relation of $x$ and $T$ is described by a triple-well potential $U(x,T)=-\frac{1}{\tau}\sum_{i=1}^3 \mathcal{N}_i (x;\mu_i,\sigma_{\text{p}i}) h_i(T)$ where $\mathcal{N}_i (x;\mu,\sigma_{\text{p}i})$ is the Gaussian distribution with mean $\mu_i$ and standard deviation $\sigma_{\text{p}i}$, $h(T)$ is a temperature dependent function. By designing $\mathcal{N}_i(x;\mu_i,\sigma_{\text{p}i})$ and $h_i(T)$, the position and the depth of the well can be controlled. The parameter of each well is 
        $\mu_1=0.16$, $\sigma_{\text{p}1}=0.2$, $h_1(T)=T^2$, 
        $\mu_2=0.55$, $\sigma_{\text{p}2}=0.11$, $h_2(T)=(1.5-T)^4$,
        $\mu_3=0.88$, $\sigma_{\text{p}3}=0.11$, $h_3(T)=1.8(1.4-T)^4$. 
    With the choice of parameters, our triple-well potential has minima at $x=\mu_1, \mu_2, \mu_3$.
    The potential minimum at $x=\mu_1$ is shallow, whereas the potential minima at $x=\mu_2,\mu_3$ are deeper.
    The global minimum at the low $T$ is $x=\mu_3$, but there is a large energy barrier between $x=\mu_2$ and $x=\mu_3$ at low $T$ so that the transition from $x=\mu_2$ to $x=\mu_3$ is unlikely.
    The shape of the potential for different temperatures is shown in Fig.~\ref{fig:3well.model}(a).
    We design the triple-well potential to imitate the disordered state in the self-assemblies of DDQC by $x=\mu_1$, and the metastable state ($\sigma \approx 0.6$) and the global minimum ($\sigma \approx 0.8$) correspond to $x=\mu_2$ and $x=\mu_3$, respectively. 

    We set $\tau=500$, $dt=1$, and $\xi_x$ following a normal distribution with mean zero and standard variation of 0.022. The parameters during the training of RL are chosen to be the same as the case for the DDQC, except that at each $T$, the number of update steps is set to 1000.  
    
\subsection{Triple-well potential result}
    
    \begin{figure}[!]
        \centering
        \includegraphics[width=0.8\textwidth]{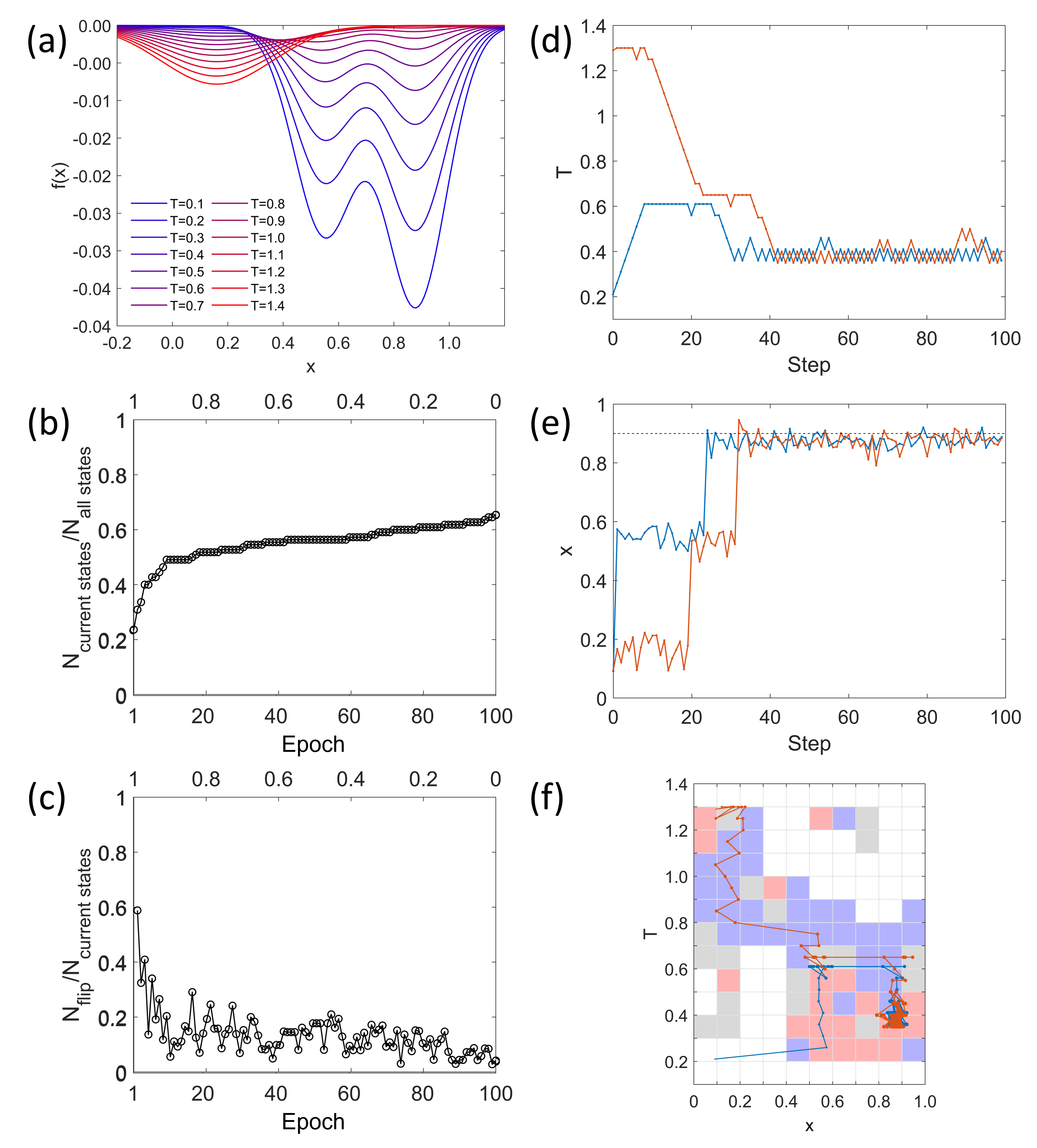}
        \caption{Q-learning for the simple model trained with random $T_0$ and number of epochs $N_\text{e}=100$. (a) The triple-well potential. (b) The ratio of the number of accessed states to total states and (c) the ratio of flipped-policy states to accessed states during training. (d-e) The states $T$ and $x$ of two independent samples and (f) their trajectories on the policy plane, noting that the point size increases with the step.}
        \label{fig:3well.model}
    \end{figure}
    To get a deeper insight into the mechanism of RL for DDQC formation, we apply Q-learning for a simple model.
    In this model, the state $x$ (analogical to the state $\sigma$ of DDQC) evolves under the triple-well potential shown in Fig.~\ref{fig:3well.model}(a). 
    We design the dependence of $x$ on the temperature analogous to that of $\sigma$ in DDQC. At high temperature, the local minimum is at $x=0.16$, similar to the low $\sigma$ structure. As the temperature decreases, this local minimum disappears, and two additional local minima appear at $x=0.55$ and $x=0.88$. The former value imitates the metastable state of a DDQC with many defects, whereas the latter corresponds to DDQC with fewer defects. By introducing the noise, a state $x$ can jump from one well to the other well under intermediate temperature $T$. 
    
    The results of the training and testing are given in Fig.~\ref{fig:3well.model}. 
    We use the number of epochs $N_\text{e} = 100$ and the random initial $T=T_0$ at the beginning of each epoch.
    Figure~\ref{fig:3well.model}(b,c) shows the ratio of the number of accessed states and the convergence of the policy during training of the Q-table.
    As the number of epochs increases (from $\epsilon=1$ to $\epsilon=0$), the number of accessed states increases, and the flip ratio converges slowly toward zero. 
    Figure~\ref{fig:3well.model}(d-f) depicts the time evolution of the states $T$ and $x$ of two testing samples and their trajectories on the policy plane. Starting with either a high or low value of $T_0$, the temperature quickly reaches $T\approx 0.6$, at which $x$ fluctuates around the middle local minimum $x=0.55$. 
    After some time, $T$ further decreases to a lower value $T \approx 0.4$, and accordingly, $x$ goes to the deepest well. 
    The estimated policy suggests two regions: decrease $T$ when $T>0.6$ and increase $T$ when $T<0.6$. The boundary between the two regions is analogical to the \UL{characteristic} temperature for the DDQC case.        

\section{\UL{Reinforcement learning for DDQC of isotropically interacting particles}}\label{secS:iso}
   
\begin{figure}[!]
        \centering
        \includegraphics[width=0.48\textwidth]{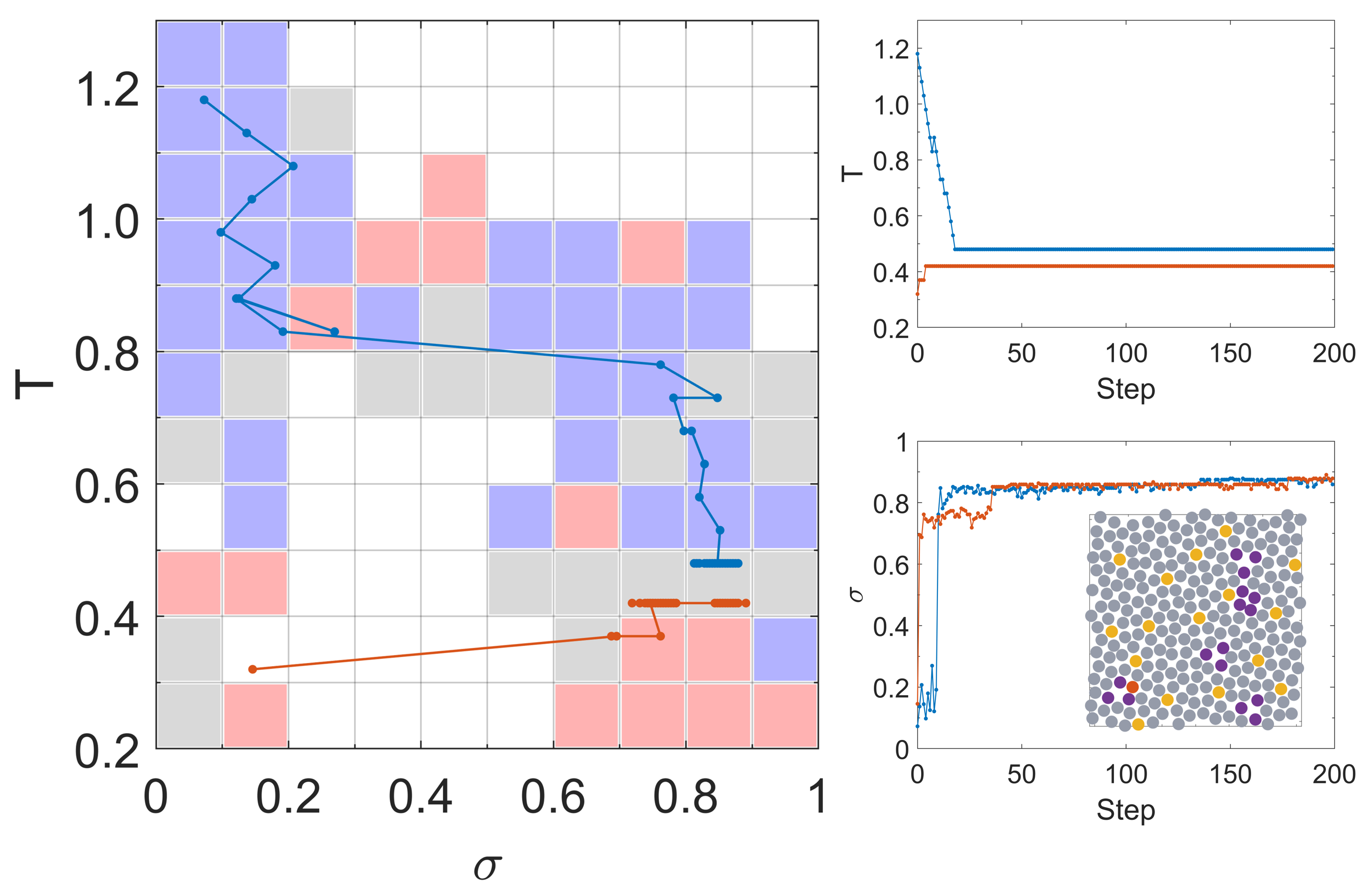}
        \caption{Reinforcement learning for DDQC of isotropically interacting particles. (Left) The trained policy with selected trajectories during tests and (right) the corresponding temperature and $\sigma$ of the tests. The mean and standard deviation of $\sigma$ from 20 independent samples are 0.84 and 0.03, respectively. The snapshot at the last point of the (orange) trajectory is given. The training condition for the isotropic particle system is the same as that of DDQC target with patchy particle system, except that the number of steps of each epoch is $N_\text{step}=100$ and the area fraction is 0.81.}
        \label{fig:Isotropic}
    \end{figure}

    In this section, we perform RL for target DDQC assembled by particles interacting with the isotropic potential. The purpose is to show the versatility of RL in handling different physical systems. As shown in Fig.~\ref{fig:Isotropic}, the trained policy for the DDQC target from particles with isotropic interactions has many 'blue' and 'grey' elements at $T \geq 0.4$, which suggests that just decrease the temperature to $T\approx 0.4$.
    This result is in contrast with that of patchy particles shown in the main text.
    For the isotropically interacting particles, simpler temperature protocol without \UL{characteristic} temperature works for the DDQC formation. 
        
    For the test starting from high initial temperature $T_0 \approx 1.2$, the temperature rapidly decreases to $T=0.5$. When the temperature passes through $T=0.8$, the particles quickly assemble into a DDQC structure whose $\sigma \approx 0.8$. The test of low initial temperature $T_0\approx 0.3$ shows that even at such a low temperature, the structure of $\sigma \approx 0.75$ can be formed immediately. Then, the policy suggests keeping $T\approx 0.4$ so that the quality of the DDQC can be improved as $\sigma>0.8$. Compared to the DDQC of patchy particles, the RL, in this case, does not feel about the existence of a \UL{characteristic} temperature $T^*$ although $\sigma$ increases drastically as $T\approx 0.8$. The agent learns through training that, for isotropic particles, the complex temperature protocol, as we have seen for the patchy particles, is not necessary to make the DDQC without defects.    
    
\section{Discussions of reinforcement learning and its hyperparameters}
\begin{figure}[!h]
        \centering
        \includegraphics[width=0.6\textwidth]{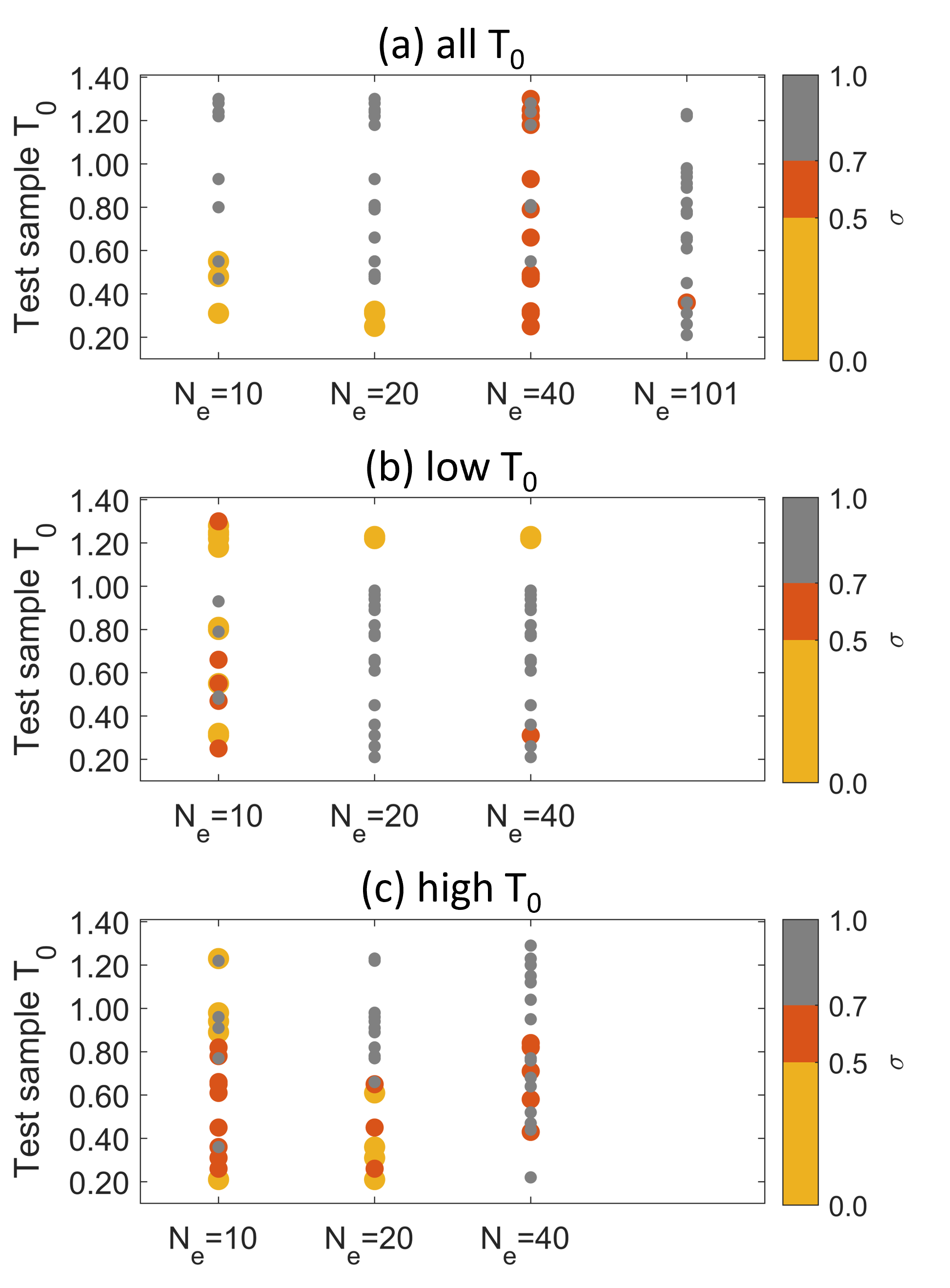}
        \caption{Performance of RL tests at different training sets at the condition (a) all $T_0$, (b) low $T_0$ and (c) high $T_0$ of each epoch in Table~\ref{table:train.DDQC.batch}. The horizontal axis labels indicate the number of epochs $N_\text{e}$ during training. The vertical axes indicate the random initial temperature $T_0$ during the tests. The colours of the spots indicate $\sigma$ values during the tests. The size of the point is for better visualisation. 
        In general, the policy is better (the test has $\sigma>0.7$) with increasing trained data. Adding constraint on $T_0$ during training can help to reduce computational cost, however the policy only works well in within the constraint.}
        \label{fig:S1prior}
        \end{figure}

    As shown in Fig.~4 in the main text, the convergence to reach the steady state of the iteration for optimising the policy is very slow. 
    Therefore, many epochs are required.
    Here, we discuss whether we may reduce the computational/training cost using prior knowledge.
    During training, instead of using random initial temperature $T_0$ for each epoch, we consider the situation of fixed $T_0$ either at high or low temperature. 
    We also vary the number of epochs for each training set. The policies of those training sets are then evaluated. The conditions and results are summarised in Table~\ref{table:train.DDQC.batch} and Fig.~\ref{fig:S1prior}. 
    \begin{table}[!h]
    \small
    \centering
        \setlength\extrarowheight{0pt} 
        \begin{tabular}{|p{5.5cm}|l|l|l|l|} 
            \hline
            \textbf{Parameters} &\textbf{All $T_0$ set }    &\textbf{Low $T_0$ set }   &\textbf{High $T_0$ set }    &\textbf{Off-policy} \\                      
            \hline        
            Target, $\sigma^*$          &0.91            &0.91            &0.91                   &0.81\\               
            Initial temperature, $T_0$               &[0.2, 1.3]     &0.22           &1.22                  &1.22   \\           
            Number of epochs, $N_\text{epoch}$         &${10, 20, 40, 101}$     &${10,20,40}$   &${10,20,40}$  &1  \\
            $\epsilon$-greedy          &decrease &  decrease         &  decrease              &1  \\
            Number of steps in each epoch, $N_\text{step}$          &  200              &  200              &200           &4000  \\        
            \hline  
        \end{tabular}    
        \caption{Setting of training sets for DDQC target by patchy particles considering the effect of the number of epochs and prior knowledge of the initial temperature $T_0$ of each epoch. The value of other hyperparameters in RL is given in Table 1 in the main text.} 
        \label{table:train.DDQC.batch}
    \end{table}

    In general, the more epochs used during training, the better the policy is; namely, more structures with $\sigma > 0.7$ can be obtained during the test.
    More importantly, the value of $T_0$ during training affects the performance of the policy. For example, even when the number of epochs is $N_\text{e}=20$, the policy of `low $T_0$' training set works well for the test with low $T_0$ but fails for the test with high $T_0$.
    On the other hand, the policy trained by the `high $T_0$' training set works well for the test with high $T_0$ but fails for the test with low $T_0$.
    In other words, if we have a constraint of the initial temperature $T_0$ in the system, we may reduce the computational cost.
        This is because the search space during training is smaller.
        The drawback of the constraint is that the trained policy cannot work outside the constraint.

\begin{figure}[!h]
        \centering
        \includegraphics[width=0.8\textwidth]{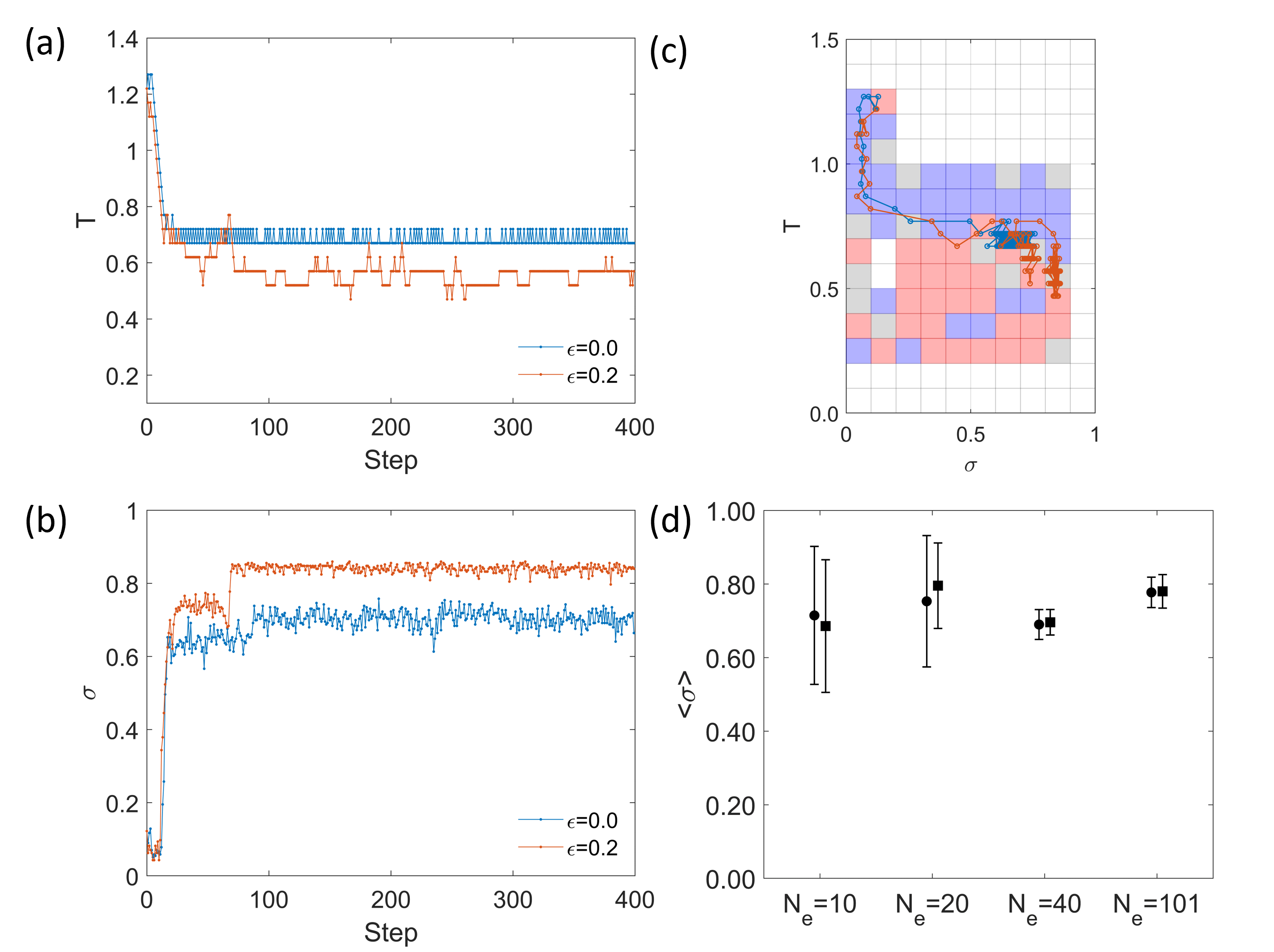}
        \caption{The effect of adding noise on testing. (a,b) Progress of $T$ and $\sigma$ at $\epsilon=0$ and $\epsilon=0.2$ of selected samples for the policy obtained in Fig.~4 and (c) corresponding trajectories in the policy plane. (d) Error bar graph showing the effect of adding noise on the performance of the testing for different training data sets (parameters are given in Table~\ref{table:train.DDQC.batch} with $\epsilon=0.0$ (filled circles) and $\epsilon=0.2$ (filled squares). 
            Sometimes adding noise can help the trajectories escape the metastable state, but the effect is statistically insignificant.}
        \label{fig:S2test.allT0.addnoise}
    \end{figure}

    In this work, we use the Q-table to perform the Q-learning.
    The DDQC self-assembly has the continuous states $\sigma$ and $T$ whose essential features can be captured even after discretisation.
    In this case, the method of Q-table works and is easily implemented. 
    However, the drawback is that the policy is sensitive to such discreteness, as it can not distinguish between the adjacent states and as a result, the agent is trapped in a state-action space and not able to find a better solution. For example, some testing samples trapped in the metastable states can escape from the metastable states and become DDQC, while others cannot escape. In order to avoid such local traps, one may consider a non-zero epsilon value (finite amplitude of noise) during testing. For example, one may set $\epsilon=0.2$ for testing so that the action on the temperature is random instead of strictly following the policy. The result in Fig.~\ref{fig:S2test.allT0.addnoise} shows that sometimes adding noise can help the test escape the metastable state and become DDQC. However, $\epsilon=0.2$ does not always result in DDQC and statistically the test results are not improved by the noise. Therefore, in the current setting, discretisation of the state space does not seem to have serious effects on the estimated policy. \\

\begin{figure}[h!]
    \centering
    \includegraphics[width=0.8\textwidth]{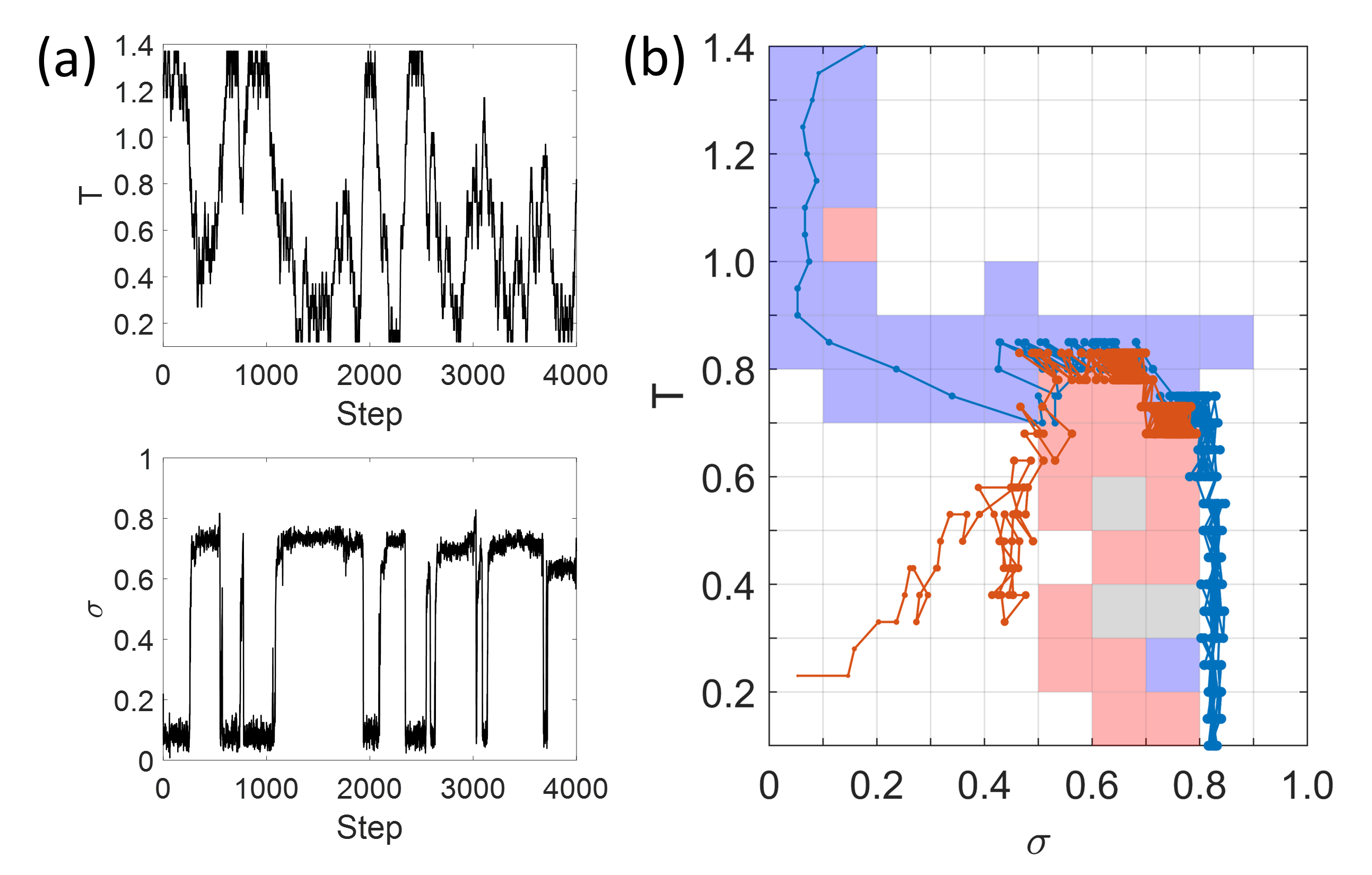}
    \caption{Off-policy training for DDQC in Table \ref{table:train.DDQC.batch}. (a) The states $T$ and $\sigma$ during training. (b) Policy and the trajectories of testing samples at different $T_0$ on the policy plane. The tests start from the left side, and the marker size increases with time. 
        Two regions, decreasing temperature (blue mesh elements) and increasing temperature (red mesh elements), can be seen on the policy plane. The test with high $T_0$ follows the policy, that is to decrease $T$ to the \UL{characteristic} temperature $T^*$, and stay there so that the DDQC structure is formed. The test with low $T_0$ starts at the inaccessible states. The action applied to the temperature is random until the trajectory hits the accessed states. Then, the trajectory follows the estimated policy and eventually DDQC structure is observed. }
            \label{fig:S3off-policy}
    \end{figure}

    So far in this study, the policy is trained with the $\epsilon$-greedy approach. The choice of the action is dependent on the current $\epsilon$ and the Q-table. Here we discuss that the RL also works even when the training is off-policy, i.e. during training, the actions at every step are randomly chosen from the action space ($\epsilon=1$). 
    The result can be found in Fig.~\ref{fig:S3off-policy}. Similar to the policy trained with $\epsilon$-greedy in Fig.~4, two regions, decreasing temperature and increasing temperature, divided by the same \UL{characteristic} temperature can be seen on the policy plane. The tests also show that the DDQC structures are observed. 
    We can see that the number of accessed states in this case is less than the one in Fig.~4. 
    This indicates that the transition probability between the accessed states is higher, and policy at the accessed states is well trained. 
    The non-accessed states have random policies. 
    In our system, these policies work because, after random fluctuations of temperature, the trajectory hits the region of accessed states and then it goes to the \UL{characteristic} temperature. Eventually, the structure becomes DDQC. 
    For this reason, the completely off-policy training works in our system. 
    Still, we suspect it does not work for other systems.

\bibliography{RLpaper}
\bibliographystyle{unsrt}